\documentclass[a4paper,fleqn,usenatbib,useAMS]{mnras}

\usepackage{color}

\usepackage{graphicx}	
\usepackage{amsmath}	
\usepackage{amssymb}	
\usepackage{multicol}        
\usepackage{bm}		
\usepackage{pdflscape}	
\usepackage{adjustbox}
\usepackage{threeparttable}
\usepackage{physics}

\title[Reconnection diffusion in different sonic regimes]{Magnetic flux transport via reconnection diffusion in different sonic regimes of interstellar MHD turbulence}

\author[C. N. Koshikumo et al.]{C. N. Koshikumo,$^{1}$\thanks{E-mail: camila.koshikumo@gmail.com (CNK)}
R. Santos-Lima,$^{1}$\thanks{E-mail: reinaldo.lima@iag.usp.br (RSL)}
M.V. del Valle,$^{1}$\thanks{E-mail: mvdelvalle@usp.br (MVdV)}
E. M. de Gouveia Dal Pino,$^{1}$\thanks{E-mail: dalpino@iag.usp.br (EMdGDP)}
\newauthor
G. Guerrero,$^{2}$\thanks{E-mail: guerrero@fisica.ufmg.br (GG)}
A. Lazarian$^{3,4}$\thanks{E-mail: lazarian@astro.wisc.edu (AL)}\\
$^{1}$Instituto de Astronomia, Geofísica e Ciências Atmosféricas, Universidade de São Paulo, SP 05508-090 \\
$^{2}$Physics Department, Universidade Federal de Minas Gerais, Av. Antonio Carlos, 6627, Belo Horizonte, MG, Brazil, 31270-901 \\
$^{3}$Department of Astronomy, University of Wisconsin, 475 North Charter Street, Madison, WI 53706, USA \\
$^{4}$Center for Computation Astrophysics, Flatiron Institute, 162 5th Ave, New York, NY 10010 }

\date{Accepted XXX. Received YYY; in original form ZZZ}

\pubyear{2025}

\begin{document}
\label{firstpage}
\pagerange{\pageref{firstpage}--\pageref{lastpage}}
\maketitle

\begin{abstract}
Turbulence and magnetic fields are components of the interstellar medium and are 
interconnected through plasma processes. In particular, the magnetic flux transport in the presence of
magneto-hydrodynamic (MHD) turbulence is an essential factor for understanding star formation.
The theory of Reconnection Diffusion (RD), based on statistics of Alfvénic turbulence,
predicts a dependence of the diffusion coefficient of the magnetic field on the Alfvénic Mach
number $M_A$.
However, this theory does not consider the effects of compressibility which are
important in the regime of supersonic MHD turbulence.
In this work, we measure the diffusion coefficient of magnetic fields in sub-Alfvénic 
MHD turbulence, with different sonic Mach numbers $M_S$.
We perform numerical simulations of forced turbulence in periodic
domains from the incompressible limit to the supersonic regime. We introduce two methods to
extract the diffusion coefficient, based on the analysis of tracer particles.
Our results confirm the RD assumption regarding the correspondence between the diffusion of
magnetic field and that of fluid Lagrangian particles. The measured diffusion rate provided
by incompressible turbulence agrees with the suppression predicted by the RD theory in the 
presence of strong magnetic fields: $D \propto M_A^3$. Our simulations also indicate an
increase in RD efficiency when the turbulence is compressible. The dependency on $M_A$ and
$M_S$ from the simulations can be described by the relation 
$D \propto M_A^{\alpha}$, where $\alpha (M_S) \approx 3 / (1 + M_S)$.
This quantitative characterization of $D$ is critical for modeling star formation in turbulent molecular
clouds and evaluating the efficiency of this transport compared to other mechanisms.
\end{abstract}




\section{Introduction}

Turbulence and magnetic fields are present in most astrophysical 
media and their important role is widely acknowledged. 
Since magnetohydrodynamic (MHD) turbulence is nearly ubiquitous, its significant
transport effects must be considered in the environment, including the transport 
of cosmic rays, material, and scalar and vector fields associated with the medium,
such as temperature (heat transport) and angular momentum.
One of the primary transport effects is the diffusion of magnetic fields, which
can challenge the assumption of the ``frozen-in'' condition for large-scale magnetic
fluxes, even when ohmic resistivity is considered negligible at these scales.
Environments of particular interest 
are the diffuse interstellar medium (ISM) and molecular clouds. 
Several pieces of observational evidence confirm that these media are both turbulent and magnetized:
power spectrum measurements of electron density in the diffuse ISM (e.g. 
\citealt{armstrong1995electron, chepurnov_lazarian_2010}), broadening of molecular lines \citep[e.g.][]{larson1981royal}, 
measurement of velocity spectra 
(e.g. \citealt{lazarian2000velocity, padoan2009power, yuen2019comment}), 
Faraday rotation statistics (\citealt{haverkorn2008outer}), power spectrum of synchrotron 
fluctuations \citep{chepurnov1999galactic}, 
measurements of the polarization of light by magnetic field-aligned interstellar dust absorption and emission (e.g. 
\citealt{andersson_etal_2015, versteeg_etal_2023, versteeg_etal_2024, angarita_etal_2023, angarita_etal_2024, yasuo_etal_2024}).
See also selected reviews describing turbulence in molecular clouds:
\citet{elmegreen2004interstellar, mac2004control, mckee2007theory}.

Therefore, turbulent magnetic flux transport is always present and occurs at a 
rate defined by the turbulence regime in the medium.
The ISM can be classified into phases according to its temperature, density, and 
ionization degree. The physical conditions of these phases, including the degree of 
magnetization, establish the sonic and Alfvénic regimes of the local turbulence.
Nonetheless, inferring these regimes accurately at sub-parsec scales and below 
(up to tens of AU) from observations can be challenging
(see however the interesting case study in \citealt{hull2017unveiling}; see also recent 
polarization studies in \citealt{angarita_etal_2023, versteeg_etal_2024, yasuo_etal_2024}). 

In the context of star formation, an efficient magnetic field transport mechanism 
is crucial for both theoretical and observational understanding of the various phases
involved in the process (e.g., \citealt{santos2010diffusion, santos2012role, santos2013disc, gonzalez2016magnetic, maury_etal_2022}).
For instance, it plays a key role in determining the distribution of cloud cores 
into subcritical (where magnetic forces prevent the gravitational collapse) 
and supercritical (where gravity dominates) regimes in turbulent molecular clouds 
(see \citealt{crutcher_2005, troland_crutcher_2008}).
Additionally, it may influence the potential transition of a core from subcritical
to supercritical (see \citealt{leao2013collapse} and references therein).
Observations of magnetic flux in T-Tauri stars also indicate the presence of a process that violates
the flux freezing condition during both gravitational contraction and accretion phases 
(see e.g. \citealt{galli_etal_2006, shu_etal_2006, santos2010diffusion}).
This is evident because the measured magnetic flux in these stars is several orders of magnitude lower
than that observed in typical supercritical cores, a phenomenon referred to as the ``magnetic flux problem''.
In these examples, where the insufficiency of ohmic diffusion is widely recognized, ambipolar 
diffusion could provide the required diffusion only under very special conditions 
(\citealt{shu_etal_2006, maury_etal_2022}).
Another example is protostellar disks, which are observed around low-mass stars
with typical diameters of $\sim$100 AU. Their formation and evolution are challenging to
explain without an efficient magnetic flux transport mechanism that allows the disk
material to decouple from part of the magnetic flux linking it to the surrounding 
envelope in the molecular cloud.
In the absence of this transport, the excess field lines are able to efficiently extract 
angular momentum from the disk via torsional Alfvén waves, preventing the disk from being 
supported by rotation (a problem known as ``magnetic braking catastrophe'', see for example 
\citealt{allen_etal_2003, galli_etal_2006, price_bate_2007, hennebelle_fromang_2008, mellon_li_2008, santos2013disc}).
Ambipolar diffusion cannot completely solve this problem (\citealt{shu_etal_2006, crutcher_etal_2009, 
krasnopolsky_etal_2010, krasnopolsky_etal_2011, li_etal_2011}), although some numerical studies 
excluding the presence of turbulence in these environments have shown the possibility of an 
amplification of the ohmic resistivity by the increase of temperature and density in the inner 
regions of the disk (\citealt{machida_etal_2007, machida_etal_2009}). 
Recognizing the inevitable presence of turbulence and its effects, it becomes completely necessary 
to understand its role in the transport of magnetic fields in all these situations.

The \textit{Reconnection Diffusion} (RD) mechanism was proposed by \cite{lazarian2005magnetic} 
to explain and quantify the transport of magnetic fields in magneto-hydrodynamic (MHD) turbulence, and is based on the 
theory of fast magnetic reconnection induced by turbulence (\citealt{lazarian1999reconnection}). 
This theory predicts that in turbulent media the reconnection happens within the time scale of 
turbulent eddy turnovers.
The RD idea was successfully tested to solve problems related to star formation: 
the ``magnetic flux problem'' (\citealt{santos2010diffusion, leao2013collapse}) and the ``magnetic 
braking catastrophe'' of protostellar disks (\citealt{santos2012role, santos2013disc, 
gonzalez2016magnetic}).
In all these instances, the diffusivity provided by turbulence was 
effective in redistributing the large-scale magnetic flux within the systems.
Additionally, 
\cite{lazarian2012magnetization} concluded from observations of molecular 
cloud cores that the mass-to-magnetic flux ratio of supercritical cores 
is consistent with magnetic flux transport via RD.

Later, 
several numerical studies on the formation of protostellar disks 
have addressed the ``magnetic braking catastrophe'' and attempted to evaluate the role of 
magnetic flux diffusion provided by RD in the formation of a rotationally sustained disk, 
in comparison to the role of other mechanisms, such as misalignment between angular momentum and magnetic 
field, ambipolar diffusion, ohmic resistivity (see e.g. \citealt{joos2012protostellar, 
joos2013influence, hennebelle2019role} and references therein).
\citet{lam_etal_2019}, for example, concluded that ambipolar diffusion and turbulence, 
with each mechanism dominating in different regions of the disk and during different 
evolutionary stages, are required to form a persistent disk.
In general, it is difficult to quantify the magnetic flux transport provided by turbulence 
in global simulations that represent complex, sometimes multi-scale, scenarios of the star formation 
process. A solid understanding of the RD mechanism and its diffusivity is required to 
plan numerical setups with sufficient resolution to represent it and to correctly 
interpret the results, as well as to understand the possible numerical artifacts that can interfere 
with the RD diffusivity. Therefore, it is vital to test the theoretical
predictions of the dependence of the RD diffusion coefficient ($\eta_{\rm rd}$) on turbulence
regime
to understand its behavior in the more realistic case of 
compressible MHD turbulence in the ISM, 
going from the subsonic to the supersonic regime. The latter case may be
particularly relevant on the scales of molecular cloud cores.

More recently, \citet{santos2021diffusion} used 3D MHD
simulations,
to test the RD  diffusion coefficient as a function of the 
\textit{Alfvénic Mach number} $M_A \equiv U/V_A$ 
(where $U$ is the characteristic turbulent velocity, $V_A = B/\sqrt{4\pi \rho}$ is 
the Alfvén speed, with $B$ representing the magnetic field strength and $\rho$ the gas density)
of the turbulence.
This study 
constituted the first attempt to simulate stationary weak MHD 
turbulence (the scenario invoked by RD theory) in the presence of finite compressibility. 
The results of this previous study 
support the predictions of the RD theory, at least in the
cases with the weakest compressibility. 

The study presented in \cite{santos2021diffusion} 
also highlighted how a common numerical setup for simulating forced turbulence in the presence 
of a strong uniform magnetic field — characterized by a relatively small ratio between the 
domain size parallel to the field and the turbulence injection scale, an isotropic distribution
of forced modes in Fourier space, and periodic boundary conditions — can introduce artificial
effects that alter the turbulence regime and, consequently, the RD transport rate of 
large-scale magnetic fields. 
When these 
artificial effects are avoided, the resulting diffusivity aligns more closely with the 
dependency predicted by the RD theory.
Moreover, the quantitative behavior of RD 
in moderately and highly compressible turbulence remains unexplored.

The aims of this work are to determine the dependence of the magnetic flux diffusion coefficient 
on the parameters of MHD turbulence through 3D direct numerical simulations.
This study numerically tests, for the first time, the RD mechanism 
in the incompressible and supersonic turbulence regimes.

This work is organized as follows. In \S2 we discuss the main assumptions and predictions 
of the RD theory to be tested. The methodology for the numerical simulations and the analysis developed 
in this study are described in \S3, and the results are presented in \S4. In \S5 we discuss our results
and our major findings are summarized in \S6.

\section{Reconnection Diffusion theory for Alfvén waves turbulence}\label{sec:RD_Theory}

The process by which the magnetic field permeating an electrically conductive flow (such as 
the astrophysical plasma of the ISM) changes its topology depends on whether the flow state 
of the fluid is laminar or turbulent. In the turbulent scenario, the chaotic gas motions produce 
conformations in the field lines that give rise to magnetic reconnection micro-sites. The rate 
of this process is independent of the 
magnetic field diffusivity provided by one or more microphysical 
mechanisms, such as ohmic resistivity, 
ambipolar diffusion, 
Hall resistivity.
Reconnection micro-sites are 
continuously formed and distributed throughout the plasma volume,
in all scales of the turbulence.
As a consequence, the 
topology of the field lines can be changed, allowing the transport of large-scale magnetic flux 
through the gas (\citealt{lazarian1999reconnection, lazarian2005magnetic, eyink2011fast,
eyink2013flux}).

The effective diffusion coefficient for the large-scale magnetic field, $\eta_{\rm rd}$, 
will depend only on the parameters 
of turbulence, and not on the details of the microphysical transport phenomena, 
provided that an inertial range for the turbulence exists.
The last condition holds when the magnetic Reynolds number, $R_M \equiv L U / \eta$, 
is greater than 1, where $L$ and $U$ are the typical length and velocity of the fluid and $\eta$ is the
magnetic diffusivity. 
The RD coefficient 
$\eta_{\rm rd}$ predicted by the theory is given by (\citealt{lazarian_2006}):
\begin{equation} \label{eta_rd}
    \eta_{\rm rd} \sim L U \min (1, M_A^3),
\end{equation}
where $U$ is the turbulent velocity at the injection scale $L$. 
When magnetic forces have minor impact on the dynamics of the turbulent cascade (the 
super-Alfvénic case, $M_A > 1$), the above expression results in the usual 
diffusion coefficient of
hydrodynamic turbulence $\sim LU$. The sub-Alfvénic case ($M_A < 1$) takes into 
account the change in the nature of turbulence due to magnetic forces (MHD turbulence).

For clarity, below we explicitly state and 
briefly describe the implicit assumptions made in the derivation of the 
relation~(\ref{eta_rd}) for the non-trivial case of sub-Alfvénic turbulence.
These assumptions are justifiable 
only for the ideal case of incompressible turbulence, for which the scaling laws are 
more easily described and are better understood from a theoretical point of view (more 
details in \citealt{eyink2011fast}; see also \citealt{lazarian1999reconnection, lazarian2005magnetic, lazarian2012turbulence}).
We tested them in our simulations with different compressibility
levels, and some of them did not prove to be valid (see \S5 for further discussions).

\textbf{It is assumed that the diffusion of magnetic flux occurs at a rate comparable to that of Lagrangian fluid particle diffusion.}
The transport rate of interest is the diffusivity transverse to the local mean magnetic field, 
described by the coefficient $\eta_{\rm rd}$. This magnetic flux diffusivity is assumed to be 
similar to the diffusion of a fluid particle in the direction perpendicular to the local mean 
magnetic field, described by the coefficient $D_\perp$. This coefficient is defined statistically by:
\begin{equation}\label{d_perp}
     D_\perp = \int_{-\infty}^{+\infty} \dd t' \langle \delta {\bf u}'_\perp (0) \cdot \delta {\bf u}'_\perp (t') \rangle,
\end{equation}
where $\delta {\bf u}'(t')$ represents the velocity fluctuations of the Lagrangian fluid 
particle at time $t'$, that is, the velocity of the fluid at the particle's position at 
time $t'$, and the \textit{brackets} $\langle \cdot \rangle$ represent a spatial average 
in a statistically homogeneous system.

\textbf{It is assumed that the phase correlation of the Alfvén waves decays exponentially in time.}
Purely incompressible MHD turbulence can be described in terms of non-linearly interacting 
Alfvén waves. When taking into account only Alfvén wave packets of a characteristic scale 
$l$ perpendicular to the local average field, and a single frequency $\omega_A$ (which 
corresponds to a characteristic length $l_\parallel$), the velocity $\delta {\bf u}' (t')$ 
is given by:
\begin{equation}
     \delta {\bf u}'_\perp (t') = {\bf \textbf{ê}_A} \delta u_l \cos \{ \omega_A t' + \phi (t') \},
\end{equation}
where ${\bf \textbf{ê}_A}$ is the direction of the velocity fluctuation of the Alfvén wave.
Therefore, the even part of the correlation function within Eq.~(\ref{d_perp}) is:
\begin{equation}
     \langle \delta {\bf u}'_{l, \perp} (0) \cdot \delta {\bf u}'_{l, \perp} (t') \rangle = \delta u^2_l \cos (\omega_A t') \langle \cos \phi_{t'} \cos \phi_0 \rangle,
\end{equation}
where we assume that the phase correlation $G(t') \equiv \langle \cos \phi_{t'} \cos \phi_0 \rangle$ 
is an even function of $t'$ (that is, $G(t' ) = G(|t'|)$). Assuming 
$G(t') = \exp(-|t'|/\tau_{\text{dec}})$ 
\citep{eyink2011fast}, 
where $\tau_\text{dec}$ is the decorrelation time, 
the integral of Eq.~(\ref{d_perp}) results in:
\begin{equation}\label{d_perp_tau_dec}
     D_\perp (l) \sim \delta u_l^2 \frac{\tau_\text{dec}}{(\omega_{A, l_\parallel} \tau_\text{dec})^2 + 1}.
\end{equation}

The above assumption of an exponential phase decorrelation will also be tested 
in the present work.

\textbf{It is assumed that the MHD turbulence develops in the weak regime for $M_A < 1$.}
The decorrelation time $\tau_\text{dec}$ of the wave phases at the injection scale is 
assumed to be of the order of the energy transfer or cascade time for smaller scales.
This cascading time is given by the weak 
MHD turbulence theory (\citealt{lazarian1999reconnection, galtier2000weak}).

The weak MHD turbulence regime (or wave turbulence) can develop if the non-linear timescale 
for the wave packets, $\sim l/\delta u_l$, is much longer than the wave crossing time 
$\sim \omega_A^{-1} \sim l_\parallel / V_A$. In this scenario, energy transfer to the 
smaller scales only occurs in the direction perpendicular to the local magnetic field, until 
$l$ becomes small enough so the two characteristic times become similar (``critical balance''), 
below which the cascade evolves in the strong regime (\citealt{goldreich1995toward}).

In the weak regime, the cascade time for balanced turbulence is 
$\tau_l \sim (l / \delta u_l) (V_A / \delta u_l)$ \citep{galtier2000weak}. 
\citet{lazarian1999reconnection}'s theory assumes that turbulence develops in the weak regime 
when the injection is sub-Alfvénic and isotropic (i.e., $L_\perp \sim L_\parallel = L$). The 
larger scale motions are the main responsible for the diffusive transport of the large scale 
magnetic flux.

If we take $\tau_\text{dec} \sim \tau_l$ with $l = L$, and assume that $\tau_L \gg \omega_A^{-1}$,
we obtain from Eq.~\ref{d_perp_tau_dec}:
\begin{equation}\label{d_perp_result}
     D_\perp (l) \sim LU M_A^3,
\end{equation}
recovering the sub-Alfvénic relation, given by Eq.~\ref{eta_rd}.

\section{Methodology}

We perform three-dimensional MHD direct numerical simulations of forced turbulence in a 
Cartesian domain with periodic boundary conditions, in the presence of a strong uniform 
background magnetic field $\mathbf{B}_0$ parallel to the $x-$direction. 
The geometry of the domain is elongated in the parallel direction (with respect to the uniform 
magnetic field), in order to avoid or minimize the effects of the limited domain size 
which compromises the validity of the classical weak turbulence regime and 
avoid the dominance of the 2D modes in the turbulence cascade
(see \citealt{nazarenko20072d} and the study of the effects of 
different domain sizes for the sub-Alfvénic turbulence in \citealt{santos2021diffusion}; 
see also \citealt{bigot_galtier_2011, alexakis_2011, alexakis_2013, lazarian_etal_2025}). 
The relation between the parallel ($x$) and perpendicular ($y,z$) sizes of the 
computational box is chosen to be $L_\parallel = 16\,L$ and $L_\perp = 1\,L$, where $L$ is the 
reference length. 

We inject turbulence in the simulations by adding Fourier components of the velocity field 
at each time step of the simulations. 
This forcing scheme excites all the velocity modes inside a spherical shell in 
$\mathbf{k}$-space (with $5 \ge k \ge 4$). The amplitudes of these modes are modulated by a 
factor $\propto \sin (2 \theta)$, where $\cos \theta = k_{\parallel} / k$. 
This distribution concentrates the energy in the velocity modes with 
$k_\parallel L / 2\pi \approx [3, 4]$ and $k_\perp L / 2 \pi \approx [3, 4]$ 
(see the `Ab' forcing scheme introduced in \citealt{santos2021diffusion}). 
The phases of the Fourier components are randomly chosen at each time step, making the forcing 
approximately delta-correlated in time.

We employ suitable codes to study each of the turbulence compressibility regimes: incompressible 
[the pseudo-spectral code \textsc{Snoopy}\footnote{https://ipag.osug.fr/~lesurg/snoopy}],
weakly 
and highly compressible [the finite volumes code \textsc{Pluto}\footnote{http://plutocode.ph.unito.it/}]. 
In order to test our methods for measuring the diffusion coefficient, we also employed the high-order finite
differences code \textsc{Pencil}\footnote{http://pencil-code.nordita.org/} in the weakly
compressible regime.
The need for the use of different codes for the aims of this work
arises from the technical difficulty of achieving the incompressible and highly compressible
limits with just one code, such as the \textsc{Pencil} code (which is conveniently
equipped with the sophisticated Test-Field method to measure the magnetic field diffusion coefficient).
Codes such as \textsc{Pluto} and \textsc{Pencil} are capable of solving
the compressible MHD equations. However, such codes does not necessarily have a readily
available and documented implementation to solve the incompressible MHD equations.
To enforce the divergence-free condition for the velocity field, it is necessary to solve a Poisson-like 
equation for the pressure, which requires special methods to be efficient in domain decomposition parallelization.
Additionally, a high-order finite difference code such as \textsc{Pencil} is not primarily designed for highly
compressible flows because the technique assumes a smooth solution
to approximate the derivatives using high-order finite differences.
For problems 
including shocks (consequence of high compressibility), 
the most obvious choice is a code such as \textsc{Pluto} code,
which integrates the equations in integral form and does not assume continuity.
The weakly compressible limit can be achieved with any compressible code by 
adopting a high sound speed relative to the flow velocity, as 
done in \citealt{santos2021diffusion}.
The computational cost to simulate this regime is high due to the small time-steps constrained by the Courant–Friedrichs–Lewy
condition (the time-step is constrained by the sound speed, which means the number of time-steps 
increases with the inverse of the compressibility of the turbulence). In this situation, 
the use of the \textsc{Pencil} code was more advantageous, as a high-order code can 
achieve faster convergence with resolution.
We also note that implementing (and validating) the Test-Field method (see Section~\ref{sec:comparison_diffusion_tracer}) 
in the \textsc{Snoopy} and \textsc{Pluto} codes is not trivial and is beyond the scope of this work.
Below, we present the set of equations evolved by each of these codes.

\subsection{Incompressible simulations}

The incompressible simulations are performed with the pseudo-spectral code \textsc{Snoopy} 
(\citealt{lesur_longaretti_2005_snoopy, lesur_longareti_2007snoopy}). 
It solves the following set of MHD equations in the Fourier space:
\begin{equation} \label{eq:velocity_incompressible}
\frac{\partial \mathbf{u}}{\partial t} + \mathbf{u}\cdot\nabla\mathbf{u} = 
-\frac{1}{\rho_0} \nabla P +
\frac{1}{4 \pi \rho_0} \left(\nabla\times\mathbf{B}\right)\times\mathbf{B}
 + \nu_3\nabla^{6}\mathbf{u} + \mathbf{f},
\end{equation}
\begin{equation}
\frac{\partial \mathbf{B}}{\partial t} = 
\nabla\times\left(\mathbf{u\times B}\right) + \eta_3\nabla^{6}\mathbf{B},
\end{equation}
where $\rho_0$ is the uniform gas density, and $P$ is the pressure, 
$\mathbf{u}$ and $\mathbf{B}$ are the velocity and magnetic fields, 
and $\mathbf{f}$ represents the bulk force per mass unit responsible for the turbulence injection. 
The divergence-free condition for both magnetic and velocity fields
($\nabla\cdot\mathbf{B} = 0$ and $\nabla \cdot \mathbf{u}=0$) during the simulation 
is imposed in the Fourier components of the fields.
$\nu_3$ and $\eta_3$ are, respectively, the coefficients of hyper-viscosity and 
magnetic hyper-diffusivity; their values were chosen after several tests to be minimal
yet sufficient to avoid the bottleneck effect at the end of the turbulent cascade.

\subsection{Weakly compressible simulations}\label{s_weak-comp-pe}

A few weakly compressible simulations are performed with the \textsc{Pencil Code} 
(\citealt{brandenburg_dobler_2002, pencil-code_collaboration_2021}), 
which uses a sixth-order finite difference spatial discretization, and solves the 
set of compressible, isothermal, MHD equations:
\begin{equation}
\frac{\partial \ln \rho}{\partial t} + \mathbf{u} \cdot \nabla \ln \rho = - \nabla \cdot \mathbf{u},
\end{equation}
\begin{equation} \label{eq:velocity}
\frac{\partial {\mathbf u}}{\partial t} + \mathbf{u} \cdot \nabla {\mathbf u} = 
- c_{s}^{2} \nabla \ln \rho + \frac{1}{\rho} \mathbf{\left( \nabla \times B \right) \times B} 
+ \nu_3 \nabla^6 \mathbf{u} + \mathbf{f},
\end{equation}
\begin{equation} \label{eq:induction}
\frac{\partial {\mathbf A}}{\partial t} + \mathbf{u} \cdot \nabla {\mathbf A} = 
\mathbf{u \times B} + \eta_3 \nabla^6 \mathbf{A}, 
\end{equation}
where $\rho$ is the density field, $c_s$ is the isothermal sound speed, 
$\mathbf{A}$ is the magnetic potential vector, 
$\mathbf{B = \nabla \times A + B_0}$ is the total magnetic field, 
$\mathbf{f}$ represents the bulk force per mass unit responsible for the turbulence injection. 
Again, $\nu_3$ and $\eta_3$ values were chosen after several tests to be minimal
yet sufficient to keep numerical stability of the simulations.

\subsection{Weakly and highly compressible simulations}
\label{sec:pluto_runs}

Simulations with different levels of compressibility 
are performed with the \textsc{Pluto} code 
(\citealt{mignone_etal_2007pluto}), 
a finite volume code with shock capture techniques for the treatment of discontinuities in 
highly compressible flows. We perform the integration of the isothermal MHD equations, 
without explicit diffusivity terms: 
\begin{equation}
\frac{\partial \rho}{\partial t} + \mathbf{\nabla} \cdot \left(\rho
		      \mathbf{u}\right)  =  0,
\end{equation}
\begin{equation}
\frac{\partial \rho \mathbf{u}}{\partial t} + \nabla\cdot\left( \rho \mathbf{u} 
	\mathbf{u} + c_s^2 \rho \right) + \mathbf{B}\times(\nabla\times\mathbf{B})  =  
\mathbf{F},
\end{equation}
\begin{equation}
\frac{\partial \mathbf{B}}{\partial t} - \mathbf{\nabla}\times (\mathbf{u}\times \mathbf{B}) = 0, 
\end{equation}
where $\mathbf{F}$ 
represents the bulk force per volume unity responsible for the turbulence injection. 
We use the second order Runge-Kutta scheme for the time evolution, and the fluxes are calculated 
employing the HLL solver and parabolic interpolation. The magnetic divergence 
is controlled 
using a divergence cleaning scheme. 

Since the divergence cleaning method can introduce artifacts in the evolution of the
magnetic field, particularly in nonlinear flows (see, e.g., \citealt{balsara_kim_2004, mocz_etal_2016}),
we repeated some of our simulations using the constrained transport (CT) scheme instead,
which ensures better control of magnetic field divergence. Both methods are implemented 
in the \textsc{Pluto} code, allowing us to compare their effects on the resulting diffusivity rates.

\subsection{Tracer particles and simulations parameters}
\label{sec:tracer_methods}

Besides the MHD fields evolved in the simulations, we integrated the evolution of $10^4$ 
tracer particles without inertia, which adopt the velocity of the fluid at their position. 
These tracers are allowed to move only in the plane perpendicular to the 
direction of the uniform magnetic field $\mathbf{B}_0$, as we are primarily interested in the 
transport properties perpendicular to 
$\mathbf{B}_0$~\footnote{The important measure for the transport of the large-scale 
magnetic field (in the direction perpendicular to the magnetic field itself) is the temporal 
correlation of the Lagrangian velocity of the fluid, supposedly tied to the field lines, 
following the waves and perturbations (according to Eq.~\ref{d_perp}). Allowing the motion of the 
particles along the magnetic fields would probably interfere with the temporal correlations 
of these ``Lagrangian points'' tied to the magnetic field, especially if the correlations 
hold for a long time.}.
The particles are initially randomly placed in the domain.
Since the boundary conditions are periodic, the number of particles is 
conserved in the domain throughout the evolution of the systems.

In order to understand the effects of the numerical resolution in our analysis we employed different 
numbers of grid points for the simulations presented: $1024 \times 64^2$ and $2048 \times 128^2$.

We indicate in Table \ref{tab:simulations_params} the important parameters of the simulations: 
the \textit{rms} velocity $v_{\rm rms}$ normalized by the reference velocity $v_0$ (although we cannot 
control directly the value of $v_{\rm rms}$ in the simulations, we regulated the turbulence injection 
in each case to keep, at the statistically stationary state, $v_{\rm rms} \approx v_0$), 
the nominal sonic Mach number $M_{S,0} = v_{0}/c_s$, 
the nominal Alfvénic Mach number $M_{A,0} = v_{0} / V_{A,0}$ (where $V_{A,0}$ is the Alfvén velocity 
calculated with the mean uniform field $B_0$ and the initial density $\rho_0$), 
the resolution, 
the time interval considered for all the analysis $[t_0, t_1]$ presented, 
and the value of the hyper-diffusion coefficients $\nu_3, \eta_3$. 

The names of the runs in Table~\ref{tab:simulations_params} include the values of  $M_{S,0}$ and
$M_{A,0}$, along with the letters `lo' or `hi' to denote the resolution ($1024 \times 64^2$ for 
`lo' and $2048 \times 128^2$ for `hi'). The final characters in the names indicate the code used:
`pe' for \textsc{Pencil}, `s' for \textsc{Snoopy}, and `pl' for \textsc{Pluto}.
All runs using the \textsc{Pluto} code use the divergence cleaning method to
control the magnetic field divergence, except for runs whose name ends in `CT',
which indicates the use of the constrained transport method (see \S~\ref{sec:pluto_runs}).

The analysis of the MHD fields presented in this work are the time average of the analysis at each 
snapshot available in the interval $[t_0, t_1]$. For all the simulations, the snapshots are spaced in time 
by 
$\Delta t = 0.76 (\ell / v_0)$, 
except for the simulations ms(...)\_ma(...)\_lo\_pe, for which the time interval is 
2$\times \Delta t$.
The position and velocity information for each of the tracer particles is recorded at time
intervals of $\delta \tau = 2 \times 10^{-3} (\ell / v_0)$ for subsequent analysis.

We should note that, in this study, we aim to connect the diffusion coefficient 
of the large scale magnetic fields (i.e., with coherence scales larger than the turbulence injection scale) 
with the MHD turbulence regimes 
characterized by the sonic and Alfvénic Mach numbers 
of the injection scale motions. Focusing on the sub-Alfvénic regime, we tried to cover the different sonic 
regimes (supersonic, subsonic, and even the incompressible limit), without assuming values or range of values 
for the sound speed and plasma beta (ratio between the thermal to magnetic pressures). All our results
are presented in dimensionless form and are applicable to interstellar regions that exhibit the
same values of $M_A$ and $M_S$ as our runs.

\begin{table*}
\centering
\caption{Simulations parameters}
\begin{threeparttable}
\begin{tabular}{l c c c c c c}
\hline\noalign{\smallskip}
\textbf{Run}		& $v_{\rm rms} / v_0$		& $M_{S,0}$\tnote{a}	& $M_{A,0}$\tnote{b}	& Resolution		& $[t_0/(\ell/v_0), t_1/(\ell/v_0)]$\tnote{c}		& $\nu_3/(\ell^5 v_0), \eta_3/(\ell^5 v_0)$\tnote{d}	\\ 
\hline\noalign{\smallskip}
ms0.1\_ma1\_lo\_pe	& 0.85				& 0.1		& 1.0		& 1024 $\times 64^2$	& $[7.6,15.2]$		& $6.2 \times 10^{-8}$	\\
ms0.1\_ma0.5\_lo\_pe	& 0.92				& 0.1		& 0.5		& 1024 $\times 64^2$	& $[7.6,15.2]$		& $6.2 \times 10^{-8}$	\\
ms0.1\_ma0.25\_lo\_pe	& 0.73				& 0.1		& 0.25		& 1024 $\times 64^2$	& $[15.2,22.9]$		& $6.2 \times 10^{-8}$	\\
ms0.1\_ma0.12\_lo\_pe	& 0.54				& 0.1		& 0.12		& 1024 $\times 64^2$	& $[22.9,38.2]$		& $6.2 \times 10^{-8}$	\\ 
\hline
incomp\_ma1\_hi\_s		& 1.11				& 0.0		& 1.0		& 2048 $\times 128^2$	& $[12.2,16.0]$		& $6.2 \times 10^{-11}$	\\
incomp\_ma0.5\_hi\_s 	& 1.08				& 0.0		& 0.5		& 2048 $\times 128^2$	& $[21.4,25.2]$		& $6.2 \times 10^{-11}$	\\
incomp\_ma0.25\_hi\_s 	& 0.86				& 0.0		& 0.25		& 2048 $\times 128^2$	& $[48.9,52.7]$		& $6.2 \times 10^{-11}$	\\
incomp\_ma0.12\_hi\_s 	& 0.76				& 0.0		& 0.12		& 2048 $\times 128^2$	& $[91.1,94.0]$		& $6.2 \times 10^{-11}$	\\
incomp\_ma1\_lo\_s		& 1.06				& 0.0		& 1.0		& 1024 $\times 64^2$	& $[11.4,15.2]$		& $6.2 \times 10^{-11}$	\\
incomp\_ma0.5\_lo\_s 	& 1.08				& 0.0		& 0.5		& 1024 $\times 64^2$	& $[11.4,15.2]$		& $6.2 \times 10^{-11}$	\\
incomp\_ma0.25\_lo\_s 	& 1.10				& 0.0		& 0.25		& 1024 $\times 64^2$	& $[26.7,30.5]$		& $6.2 \times 10^{-11}$	\\
incomp\_ma0.12\_lo\_s 	& 0.93				& 0.0		& 0.12		& 1024 $\times 64^2$	& $[87.9,91.7]$		& $6.2 \times 10^{-11}$	\\
\hline
ms0.1\_ma1\_hi\_pl	& 1.01				& 0.1		& 1.0		& 2048 $\times 128^2$	& $[7.6,15.2]$		& -			\\
ms0.1\_ma0.5\_hi\_pl	& 1.22				& 0.1		& 0.5		& 2048 $\times 128^2$	& $[7.6,15.2]$		& -			\\
ms0.1\_ma0.25\_hi\_pl	& 1.32				& 0.1		& 0.25		& 2048 $\times 128^2$	& $[7.6,15.2]$		& -			\\
ms0.1\_ma0.12\_hi\_pl	& 1.35				& 0.1		& 0.12		& 2048 $\times 128^2$	& $[7.6,15.2]$		& -			\\
ms0.1\_ma1\_lo\_pl	& 1.13				& 0.1		& 1.0		& 1024 $\times 64^2$	& $[7.6,15.2]$		& -			\\
ms0.1\_ma0.5\_lo\_pl	& 1.33				& 0.1		& 0.5		& 1024 $\times 64^2$	& $[7.6,15.2]$		& -			\\
ms0.1\_ma0.5\_lo\_pl\_CT 
& 1.34				
& 0.1		
& 0.5		
& 1024 $\times 64^2$	
& $[7.6,15.2]$		
& -			
\\
ms0.1\_ma0.25\_lo\_pl	& 1.52				& 0.1		& 0.25		& 1024 $\times 64^2$	& $[7.6,15.2]$		& -			\\
ms0.1\_ma0.25\_lo\_pl\_CT 
& 1.49				
& 0.1		
& 0.25	
& 1024 $\times 64^2$	
& $[7.6,15.2]$		
& -			
\\
ms0.1\_ma0.12\_lo\_pl	& 1.45				& 0.1		& 0.12		& 1024 $\times 64^2$	& $[7.6,15.2]$		& -			\\
\hline
ms1\_ma1\_hi\_pl	& 1.10				& 1.0		& 1.0		& 2048 $\times 128^2$	& $[3.0,15.2]$		& -			\\
ms1\_ma0.5\_hi\_pl	& 1.05				& 1.0		& 0.5		& 2048 $\times 128^2$	& $[3.0,15.2]$		& -			\\
ms1\_ma0.25\_hi\_pl	& 1.00				& 1.0		& 0.25		& 2048 $\times 128^2$	& $[3.0,15.2]$		& -			\\
ms1\_ma0.12\_hi\_pl	& 1.47				& 1.0		& 0.12		& 2048 $\times 128^2$	& $[3.0,9.2]$		& -			\\
ms1\_ma1\_lo\_pl	& 1.25				& 1.0		& 1.0		& 1024 $\times 64^2$	& $[3.0,15.2]$		& -			\\
ms1\_ma0.5\_lo\_pl	& 1.16				& 1.0		& 0.5		& 1024 $\times 64^2$	& $[3.0,15.2]$		& -			\\
ms1\_ma0.5\_lo\_pl\_CT 
& 1.23				
& 1.0		
& 0.5		
& 1024 $\times 64^2$	
& $[3.0,15.2]$		
& -			
\\
ms1\_ma0.25\_lo\_pl	& 1.07				& 1.0		& 0.25		& 1024 $\times 64^2$	& $[3.0,15.2]$		& -			\\
ms1\_ma0.25\_lo\_pl\_CT 
& 1.16				
& 1.0		
& 0.25		
& 1024 $\times 64^2$	
& $[3.0,15.2]$		
& -			
\\
ms1\_ma0.12\_lo\_pl	& 1.55				& 1.0		& 0.12		& 1024 $\times 64^2$	& $[3.0,15.2]$		& -			\\
\hline
ms3\_ma1\_hi\_pl	& 1.34				& 3.0		& 1.0		& 2048 $\times 128^2$	& $[3.0,6.9]$		& -			\\
ms3\_ma0.5\_hi\_pl	& 1.17				& 3.0		& 0.5		& 2048 $\times 128^2$	& $[3.0,6.9]$		& -			\\
ms3\_ma0.25\_hi\_pl	& 0.92				& 3.0		& 0.25		& 2048 $\times 128^2$	& $[3.0,12.2]$		& -			\\
ms3\_ma0.12\_hi\_pl	& 0.99				& 3.0		& 0.12		& 2048 $\times 128^2$	& $[3.0,6.9]$		& -			\\
ms3\_ma1\_lo\_pl	& 1.16				& 3.0		& 1.0		& 1024 $\times 64^2$	& $[3.0,15.2]$		& -			\\
ms3\_ma0.5\_lo\_pl	& 0.95				& 3.0		& 0.5		& 1024 $\times 64^2$	& $[3.0,15.2]$		& -			\\
ms3\_ma0.5\_lo\_pl\_CT	
& 0.98				
& 3.0		
& 0.5		
& 1024 $\times 64^2$
& $[3.0,15.2]$		
& -			
\\
ms3\_ma0.25\_lo\_pl	& 0.72				& 3.0		& 0.25		& 1024 $\times 64^2$	& $[3.0,15.2]$		& -			\\
ms3\_ma0.25\_lo\_pl\_CT	
& 0.75				
& 3.0		
& 0.25		
& 1024 $\times 64^2$	
& $[3.0,15.2]$		
& -			
\\
ms3\_ma0.12\_lo\_pl	& 1.13				& 3.0		& 0.12		& 1024 $\times 64^2$	& $[3.0,15.2]$		& -			\\
\hline
\end{tabular}
\begin{tablenotes}
\vskip 0.1in
\item[a] $M_{S,0} \equiv v_0 / c_s$ is the approximate sonic Mach number of the simulations.
\item[b] $M_{A,0} \equiv v_0 / v_{A,0}$ is the approximate Alfvénic Mach number of the simulations, where $v_{A,0} = B_{0} / \sqrt{4 \pi \rho_{0}}$.
\item[c] $[t_0, t_1]$ is the time interval used for the analysis.
\item[d] $\nu_3$, $\eta_3$ are the hyper-viscosity and hyper-resistivity coefficients.
\end{tablenotes}
\end{threeparttable}
\label{tab:simulations_params}
\end{table*}

\subsection{Energy and transfer spectra}

Although our simulations cover from the incompressible up to the highly compressible regimes, for simplicity 
all the spectral analysis are based on the definitions usually
applied to the incompressible turbulence limit, and based on the spectral energy density $S_{\rm 3D}(\mathbf{k})$:
\begin{equation}
	S_{\rm 3D}(\mathbf{k}) = \frac{1}{2} \bar{\rho}	\mathbf{u_{k}^*} \cdot \mathbf{u_{k}} + \frac{1}{8 \pi} \mathbf{B_{k}^*} \cdot \mathbf{B_{k}},
\end{equation}
where
$\bar{\rho}$ is the mean density,
$\mathbf{u}_\mathbf{k} = \mathcal{F}_{\mathbf k} \left\{ \mathbf{u} \right\}$ and 
$\mathbf{B}_\mathbf{k} = \mathcal{F}_{\mathbf k} \left\{ \mathbf{B} \right\}$ are the 
$\mathbf{k}$-component of the 
discrete Fourier transform of the fields $\mathbf{u}$ and $\mathbf{B}$,
respectively, and the superscript $^*$ means the complex conjugate. 
The vector $\mathbf{k}$ can be decomposed in its parallel and perpendicular 
components with respect to the uniform field $\mathbf{B}_0$: 
$\mathbf{k} = (\mathbf{k}_{\parallel} + \mathbf{k_{\perp}})$, where 
$\mathbf{k}_{\parallel} = (\mathbf{k \cdot B}_0) \mathbf{B}_0 / B_0^2$.
                                
Following \cite{santos2021diffusion}, we define the 2D power spectrum of the turbulence $E_{\rm 2D}(k_{\parallel},k_{\perp})$ 
in terms of the spectral energy density $S_{\rm 3D}(\mathbf{k})$ as
\begin{equation}
	E_{\rm 2D}(k_{\parallel},k_{\perp}) = \sum_{\mathbf{k}'} S_{\rm 3D}(\mathbf{k'}),
\end{equation}
where $k_{\parallel,\perp} = |\mathbf{k}_{\parallel,\perp}|$,
and the sum extends for all the discrete modes $\mathbf{k'}$
with components in the interval $k_{\parallel} \le | \mathbf{k}_{\parallel}' | < (k_{\parallel} + 1)$,
and $k_{\perp} \le | \mathbf{k_{\perp}'} | < (k_{\perp} + 1)$. 

From $E_{\rm 2D}(k_{\parallel},k_{\perp})$ we derive the perpendicular 
1D power spectrum $E_{\rm 1D}(k_{\perp})$:
\begin{equation}\label{eq:e1d}
	E_{\rm 1D}(k_{\perp}) = \sum_{k_{\parallel}=0}^{k_{\parallel,\max}} E_{\rm 2D}(k_{\parallel},k_{\perp}),
\end{equation}
and the total turbulent energy in the system is given by
\begin{equation}
	E_{\rm turb} = \frac{1}{2} \bar{\rho} \langle \left( \delta \mathbf{u} \right)^2 \rangle + \frac{1}{8 \pi} \langle \left( \delta \mathbf{B} \right)^2 \rangle = 
 \sum_{k_{\perp}=1}^{k_{\perp,\max}} E(k_{\perp}),
\end{equation}
with the angular brackets representing the average in space.

In order to characterize the strength of the 2D modes of the velocity field (those with 
$k_{\parallel} = 0$) that are solenoidal, we also define a perpendicular 1D spectrum 
for these modes as follows:
\begin{equation}\label{eq:spectrum_vel2d}
	E_{\mathbf{u}^{\rm 2D}_{\rm sol}} (k_\perp) = \sum_{\mathbf{k'}} \frac{1}{2} \bar{\rho}	\mathbf{u_{k, {\rm sol}}^*} \cdot \mathbf{u_{k, {\rm sol}}},
\end{equation}
where the sum extends over all the modes 
$\mathbf{k'}$
with $k_{\parallel}' = 0$ and perpendicular components
in the interval  
$k_{\perp} \le | \mathbf{k_{\perp}'} | < (k_{\perp} + 1)$, and 
$\mathbf{u_{k, {\rm sol}}} = [ \mathbf{u_{k}} - \mathbf{(u_{k} \cdot k) k} / k^2 ]$.

For the incompressible limit of the MHD equations (see for example 
Eq.~\ref{eq:velocity_incompressible}), the time derivative of $S_{\rm 3D}(\mathbf{k})$ 
inside the inertial range is given
by $T_{\rm 3D}(\mathbf{k})$:
\begin{eqnarray}
	T_{\rm 3D}(\mathbf{k}) &=& 
	\Re \left\{ \bar{\rho} \mathbf{u^*_k} \cdot \mathcal{F}_{\mathbf k} \left[ \mathbf{ \left( u \cdot \nabla \right) u } 
	- \frac{1}{4 \pi} \mathbf{ \left( \nabla \times B \right) \times B } \right] \right\} \nonumber \\
	&& - 
\frac{1}{4 \pi} 
\Re \left\{ \mathbf{B^*_k} \cdot \mathcal{F}_{\mathbf k} \left[ \mathbf{ \nabla \times \left( u \times B \right) } \right] \right\}, 
\end{eqnarray}
where $\Re \{ ... \}$ represents the real part of the complex quantity inside the braces, 
$\mathcal{F}_{\mathbf k} [...]$ symbolizes the 
$\mathbf{k}$-component of the discrete Fourier transform of the field inside the brackets,
and the dissipation and forcing terms are ignored 
as they are important only for modes $\mathbf{k}$ outside the inertial range. 
We define the perpendicular transfer spectrum $T_{\rm 1D}(k_{\perp})$ as a sum
of $T_{\rm 3D}(\mathbf{k})$ over a volume in the Fourier space:
\begin{equation}\label{eq:t1d}
	T_{\rm 1D}(k_{\perp}) = \sum_{\mathbf{k'}} T_{\rm 3D}(\mathbf{k'}),
\end{equation}
with the sum extending over all the modes 
$\mathbf{k'}$
with perpendicular components
in the interval  
$0 \le | \mathbf{k_{\perp}'} | < (k_{\perp} + 1)$.
With the above definitions, $T_{\rm 1D}(k_{\perp})$ is expected to be 
maximum and constant inside the inertial range. Positive values mean that the energy is 
flowing from larger to smaller scales, while negative values indicate an inverse cascade. 
Therefore, we define the turbulence energy transfer rate as the maximum value 
of $T_{\rm 1D}(k_{\perp})$,
\begin{equation}
	T_{\rm turb} = \max \left\{ T_{\rm 1D}(k_{\perp}) \right\}.
\end{equation}

\subsection{Statistics of tracer particles}\label{sec:statistics_tracer}

We quantify the perpendicular diffusion coefficient of the magnetic field employing two new methods 
introduced in this study, based on the statistical analysis of Lagrangian tracer particles. 

In the first method, we start calculating the discrete auto-correlation $A(\tau_k)$ in time of the 
particles perpendicular velocity:
\begin{equation}
A(\tau_k) = \frac{1}{N(\tau_k)} \sum_{i,j} \left\langle \mathbf{v}_{\perp}(t_i) \cdot \mathbf{v}_{\perp}(t_j) \right\rangle_p,
\end{equation}
where $\langle \cdot \rangle_p$ means the average
over the $N_p$ particles.
The sum is over all the $N(\tau_k)$ time pair combinations $i, j$ available,
such that $\tau_k \le (t_j - t_i) < \tau_k + \delta \tau$, where $\tau_k = k \delta \tau$
and $k = 0, 1, 2, ..., k_{\max}$. The value of $k_{\max}$ is limited by the analysis time period
(indicated in Table~\ref{tab:simulations_params})
and $\delta \tau$ represents the time interval between consecutive outputs of particle data.
The standard deviation $\sigma_A (\tau_k)$ and the standard deviation of the mean 
$\sigma_{A, {\rm mean}} (\tau_k)$ are calculated using the standard expressions, as follows:
\begin{equation}
\sigma_A (\tau_k) = \sqrt{ \frac{1}{\left\{ N(\tau_k) - 1 \right\}} \sum_{i,j} \left\{ \left\langle \mathbf{v}_{\perp}(t_i) \cdot \mathbf{v}_{\perp}(t_j) \right\rangle_p - A(\tau_k) \right\}^2 },
\end{equation}
\begin{equation}
\sigma_{A, {\rm mean}} (\tau_k) = \frac{1}{\sqrt{N(\tau_k)}} \sigma_A(\tau_k).
\end{equation}
We then integrate the discrete 
auto-correlation function $A(\tau_k)$ to estimate the perpendicular particle diffusion coefficient:
\begin{equation}\label{eq:dcorr}
D_{\rm corr} = \delta \tau \sum_{k = 0}^{k_{\max}} A(\tau_k).
\end{equation}
The diffusion coefficient obtained through this process is identified by the subscript 
\textit{corr}. 

To obtain the statistical standard deviation $\sigma_{D_{\rm corr}}$, 
we proceed as follows. 
First, we generate an auto-correlation function with noise:
$A_{\rm noisy}^{(m)} (\tau_k) = A (\tau_k) + \psi^{(m)} (\tau_k)$, where $\psi^{(m)} (\tau_k)$
is a random variable sourced from a normal distribution with variance
$\sigma^2 = \sigma_A^2 (\tau_k)$. This calculation is then repeated $N_{\rm noisy}$
times, that is, $m = 1, 2, ..., N_{\rm noisy}$. 
Next, we compute the statistical standard deviation of the $N_{\rm noisy}$ cumulants
$B^{(m)}(\tau_k) = \delta \tau \sum_{k' = 0}^{k} A_{\rm noisy}^{(m)}(\tau_{k'})$,
denoted as $\sigma_{B}(\tau_k)$. Finally, we obtain
$\sigma_{D_{\rm corr}} = \sigma_{B}(\tau_{k_{\max}})$. We
set $N_{\rm noisy} = 10,000$.

We modeled the auto-correlation function $A(\tau)$ using
\begin{equation}\label{eq:fit}
    f(\tau) = A_0 \cdot \cos \left( \omega \tau \right) \cdot \exp \left\{ - \left( \gamma \tau \right)^{\zeta} \right\},
\end{equation}
where $A_0$ (amplitude), $\omega$ (oscillatory frequency), and $\gamma^{-1}$ (decorrelation time) are the fitting parameters.
We considered both cases $\zeta = 1$ (exponential decorrelation) and $\zeta = 2$ (Gaussian decorrelation). 
In practice, we look for the parameters which give the best fit to the cumulant of $A(\tau_k)$. For this, 
during the fitting procedure we use the numerically calculated cumulant of $A(\tau_k)$ and the analytical 
function for the cumulant of $f(\tau)$.
The perpendicular diffusion coefficient obtained from the integral of the fitted function $f(\tau)$ 
will be identified by the subscript \textit{corr-fit}:
\begin{equation}\label{eq:dfit}
D_{\rm corr-fit} = \int_{0}^{\tau_{\max}} {\rm d}\tau f(\tau).
\end{equation}

Through this first method, based on the analysis of the time series of the particles' velocities,
we are able to access the auto-correlation function of the velocity of the oscillating points 
connected to the field lines, and to connect this statistics with those associated with the 
turbulence waves. However, the disadvantage of this method is the need for a high-frequency
time series of the particles' velocity to capture the fastest oscillations in the system.  

The second method is based on the direct calculation of the average quadratic displacement of the particles
on the perpendicular plane to the mean magnetic field:

\begin{equation}\label{eq:dyz}
D_{yz} = \frac{1}{N_i} \sum_{i} \left\langle \frac{1}{2 \Delta t} \left[ \mathbf{r}^2 (t_{i + 1}) - \mathbf{r}^2 (t_i) \right] \right\rangle_p,
\end{equation}
where, again, $\langle \cdot \rangle_p$ denotes the average over the $N_p$ particles,
and the sum extends over all available snapshots $i$, with $\Delta t$ representing the
time interval between consecutive snapshots, and $\mathbf{r}_n^2 (t_i)$ the quadratic 
displacement of particle $n$ at $t_i$ since time $t = 0$.
The subscript \textit{yz} identifies the diffusion coefficients calculated by this second method.
The standard deviation $\sigma_{D_{yz}}$ and the standard deviation of the mean $\sigma_{D_{yz},
{\rm mean}}$ are also calculated using the standard expressions, as follows:
\begin{equation}
\sigma_{D_{yz}} = \sqrt{ \frac{1}{\left\{ N_i - 1 \right\}} \sum_{i} \left\{ \left\langle \frac{1}{2 \Delta t} \left[ \mathbf{r}^2 (t_{i + 1}) - \mathbf{r}^2 (t_i) \right] \right\rangle_p - D_{yz} \right\}^2 },
\end{equation}
\begin{equation}
\sigma_{D_{yz}, {\rm mean}} = \frac{1}{\sqrt{N_i}} \sigma_{D_{yz}}.
\end{equation}

Compared to the first method, this second one has the advantage that it directly calculates
the diffusion of the Lagrangian points connected to the field lines, without any intermediate
calculation susceptible to systematic errors. Besides, it does not require a high-frequency
time series for the particles' positions, as it does not intend to follow their oscillatory movements.

\section{Results}

\begin{figure*}
\begin{tabular}{c}
    \input{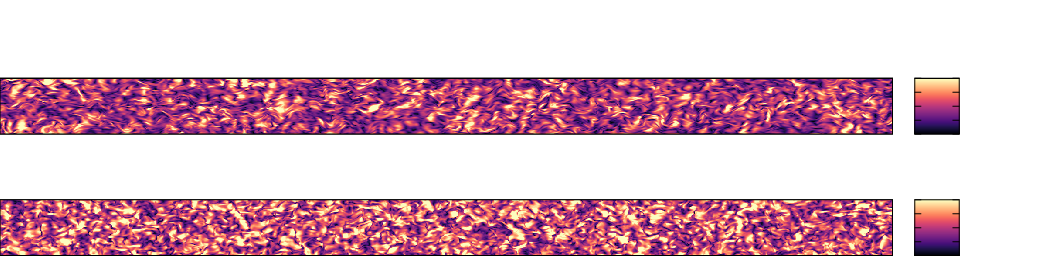} \\
\end{tabular}
\caption{Velocity amplitude distribution in the central $xy$-plane of the domain at the final time 
of the simulations. The sonic Mach number $M_S \equiv v_0 / c_s$ and the Alfvénic Mach number 
$M_A \equiv v_0/v_{A,0}$ are contained in the name of the run, shown at the top of each map. 
See Table~\ref{tab:simulations_params} for the complete description of the simulations 
parameters.}
\label{fig:maps_vrms}	
\end{figure*}

The geometry of the computational domain and the turbulence structures in the statistically 
steady state of the numerical simulations are illustrated in Figure~\ref{fig:maps_vrms} 
for 
two 
of the models presented in Table~\ref{tab:simulations_params}. 
Each map shows the velocity modulus distribution normalized by the reference value $v_0$, in the central 
$yz$-plane of the domain of the last available snapshot. 
The upper map is from an incompressible simulation, while
the bottom map corresponds to a supersonic simulations ($M_S = 3$), both with
the same value of $M_A$.

\begin{figure*}
\begin{tabular}{c}
    \input{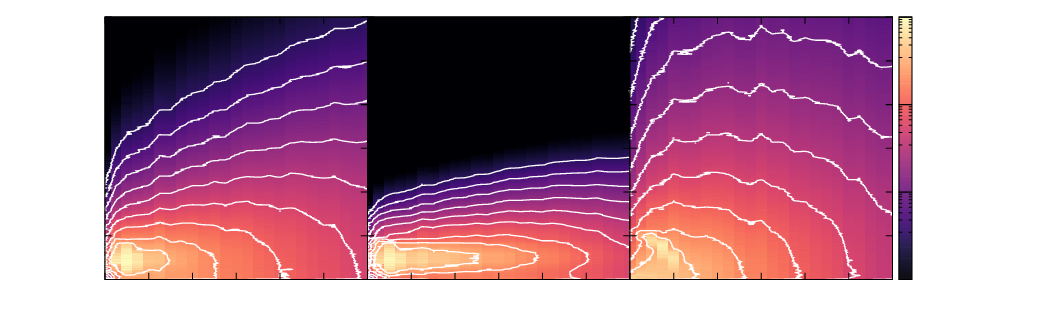}
\end{tabular}		
\caption{Two-dimensional energy spectrum $E_{\rm 2D}(k_\parallel, k_\perp)$ for simulations 
with different sonic Mach numbers: 
incompressible (leftmost and central panel) and $M_S=3$ (rightmost panel).
Each incompressible case corresponds to a different 
nominal Alfvénic Mach number $M_A \equiv v_0/v_{A,0}$: $M_A=0.5$ 
(leftmost panel), and 
$M_A=0.25$ 
(central panel).
See Table~\ref{tab:simulations_params} for the complete description 
of the simulations parameters.}
\label{fig:maps_ps}
\end{figure*}

The 2D turbulence spectra $E_{2D}(k_{\parallel},k_{\perp})$ for 
a selection of the models presented in Table~\ref{tab:simulations_params}, 
including the same models shown in Figure~\ref{fig:maps_vrms},  
are represented in Figure~\ref{fig:maps_ps}. 
These spectra are averaged over the analysis period of each model (see 
Table~\ref{tab:simulations_params}), in the statistically steady state of turbulence. 
In all spectra in Figure~\ref{fig:maps_ps}, anisotropy is evident (noticeable in the shape of the level 
curves). In the shell where energy is injected through the velocity field ($4 \le |\mathbf{k}| \le 5$) 
the modulation in amplitude (which depends on the direction) can be visualized. 
Although no energy is injected into the 2D modes ($k_{\parallel}=0$), the distribution in all the cases 
develops smoothly in this limit for any value of $k_{\perp}$. The suppression of the energy transfer 
in the direction parallel to $\mathbf{B}_0$ (the vertical direction in each plot) at scales close 
to the injection is more visible in the incompressible case (left panels), 
and it becomes less pronounced in the compressible simulations 
($M_S=3$ in the rightmost panel). The suppression of the energy transfer in the parallel direction 
(up to the transition scale $\ell_{\rm tr} \sim \ell M_A^2$ where the critical balance is achieved) 
is a prediction of the Alfvénic weak turbulence theory, and this suppression was also identified in 
the weakly compressible simulations studied in \citet{santos2021diffusion}.

\subsection{Comparison between the diffusion coefficients of the magnetic field and of the tracer particles}\label{sec:comparison_diffusion_tracer}

\begin{figure}
\input{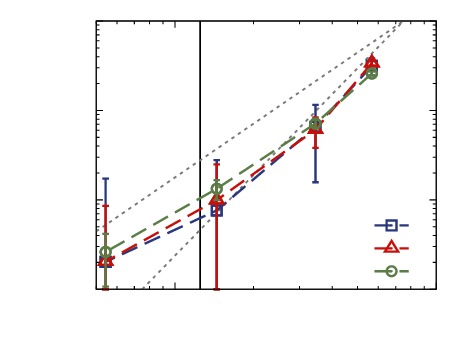}
\caption{Magnetic diffusion coefficient measured by the Test-Field $\eta_{\rm tf}$ and the 
perpendicular diffusion coefficient of the tracer particles $D_{\perp}$ as a function of 
the Alfvénic Mach number $M_A = v_{\rm rms}/ \langle v_{A} \rangle$ for the set of simulations 
ms0.1\_ma(...)\_lo\_pe, performed with the \textsc{Pencil Code}  (see Table~\ref{tab:simulations_params}). 
Each point corresponds to one run in that set. The particles diffusion 
coefficients are measured using the velocity auto-correlation function ($D_{\rm corr}$) and
the evolution of the particles displacements ($D_{\rm yz}$).
The vertical bars indicate the standard deviation in the temporal distribution 
of all quantities and do not represent statistical errors of the mean values. 
The vertical line indicates the theoretical limit of $M_A$ below which finite domain size 
effects can affect the turbulence regime (see text for more details).}
\label{fig:ma_eta_codes}
\end{figure}

The \textit{Test-Field} method 
(\citealt{brandenburg_subramanian_2005, schrinner_etal_2007, brandenburg_etal_2010}) 
has been extensively employed to extract the effective values 
of the terms in the mean-field theory describing the evolution of the mean magnetic field in 
dynamo and turbulence studies 
(e.g. \citealt{brandenburg_etal_2008, sharanya_etal_2008, karak_etal_2014, brandenburg_etal_2017, santos2021diffusion, brandenburg_etal_2023}).
We are interested here in the perpendicular diffusion coefficient of the magnetic field. 
The Test-Field method is sophisticated and precise, however until this date it is implemented in 
just one MHD code publicly available and with a wide range of applications: the 
\textsc{Pencil Code}
(see Section~\ref{s_weak-comp-pe} for more details). 
Figure~\ref{fig:ma_eta_codes} shows the diffusion coefficients extracted 
via the Test-Field method, $\eta_{\rm tf}$, for the simulation set ms0.1\_ma(...)\_lo\_pe 
(see Table~\ref{tab:simulations_params})
as a function of the Alfvénic Mach number $M_A$ ($= v_{\rm rms} / \langle v_A \rangle$). 
The perpendicular diffusion coefficients of the tracer particles, $D_{\perp}$, for the same 
simulations are also displayed, i.e. $D_{\rm corr}$ (see Eq.~\ref{eq:dcorr}) and $D_{\rm yz}$ 
(see Eq.~\ref{eq:dyz}), obtained using the two methods 
described in \S~\ref{sec:statistics_tracer}. We observe that $\eta_{\rm tf}$, $D_{\rm corr}$, 
and $D_{\rm yz}$ are all in close agreement. This result verifies the assumption of RD theory 
that the diffusion of magnetic fields occurs at a rate similar to that of the perpendicular 
diffusion of Lagrangian fluid particles 
(\S~\ref{sec:RD_Theory}).
The vertical bars shown in the Figure~\ref{fig:ma_eta_codes} 
(and in all Figures throughout this Section, unless explicitly stated otherwise) 
indicate the standard deviation in 
the temporal distribution of the quantities, and their calculation for 
$D_{\rm corr}$ and $D_{\rm yz}$ are explained in 
\S~\ref{sec:statistics_tracer}; the statistical errors of the means 
(the standard deviation of the mean in the temporal distribution) have values smaller 
than the point sizes and are not shown. 
This result also validates the two new methods introduced in this work for the analysis 
of magnetic diffusion coefficient through turbulence. Therefore, we will now apply these 
new methods to simulations that employ the \textsc{Snoopy} and \textsc{Pluto} codes, which do 
not have the Test-Field method implemented, but which are suitable for the limits of 
incompressible and supersonic turbulence, respectively.

We also observe in Figure~\ref{fig:ma_eta_codes} that, except for the simulation with the 
lowest value of $M_A$ (which is below the theoretical limit of $M_A$,
where effects due to the finite domain size can occur [see \citealt{santos2021diffusion} 
and \citealt{nazarenko20072d}
for more details]), the dependence of the diffusion coefficient 
on $M_A$ follows a power law $M_A^{\alpha}$, with $\alpha$ between 2 and 3 
(closer to $\alpha = 3$). This result coincides with those reported in 
\citealt{santos2021diffusion} for weakly compressible turbulence.

\subsection{Diffusion coefficient in the incompressible limit}

\begin{figure*}
\begin{tabular}{c}
        \input{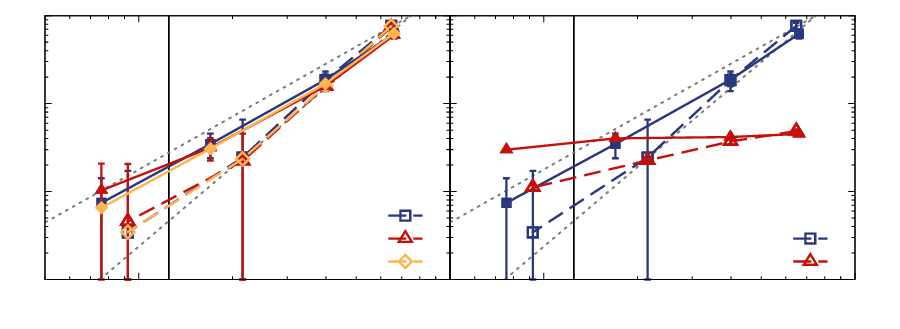} \\
\end{tabular}
\caption{Normalized diffusion coefficient of the tracer particles $D_{\perp}$ (left and right 
panels), and the rms value of the 2D component of the solenoidal velocity 
$\langle v_{2D, {\rm sol}} \rangle$ normalized by the total rms velocity $v_{\rm rms}$ (right only),
as a function of the Alfvénic Mach number $M_A = v_{\rm rms}/ \langle v_{A} \rangle$. 
Each point in the curves corresponds to an incompressible simulation in the run set performed 
with the \textsc{Snoopy} code (see Table~\ref{tab:simulations_params}). 
The diffusion coefficients are measured using the velocity auto-correlation function 
($D_{\rm corr}$), the evolution of the particles displacements ($D_{\rm yz}$), 
or calculated using the fitted curve for the velocity auto-correlation function 
($D_{\rm corr-fit}$). 
The vertical bars indicate the standard deviation in the temporal distribution 
of the quantities $D_{\rm corr}$, $D_{\rm yz}$, and $\langle v_{2D, {\rm sol}} \rangle$, 
and do not represent statistical errors of the mean values. 
The vertical line indicates the theoretical limit of $M_A$ below which finite domain size 
effects can affect the turbulence regime (see text for more details). 
Simulations with different resolutions are compared: $2048 \times 128^2$ (continuous lines) 
and $1024 \times 64^2$ (dashed lines). 
}
\label{fig:ma_eta_incomp}
\end{figure*}

Figure~\ref{fig:ma_eta_incomp} shows the dependence of the diffusion coefficient on $M_A$ for 
the incompressible simulations in Table~\ref{tab:simulations_params}. The solid lines 
connect points of the simulations with higher resolution, 
while the dashed lines connect the points of the lower resolution simulations. 
The left panel of Figure~\ref{fig:ma_eta_incomp} compares the diffusion coefficient calculated 
using the velocity correlation $D_{\rm corr}$ 
(see Eq.~\ref{eq:dcorr})
with that
using the particles spatial displacements 
$D_{\rm yz}$ (see Eq.~\ref{eq:dyz}).  
Additionally, the curve $D_{\rm corr-fit}$ (see Eq.~\ref{eq:dfit}) shows the diffusion coefficients 
obtained through the fittings of the velocity auto-correlation $A(\tau)$ for each simulation. 
Because we seek the best fit for the cumulant, in later 
times $\tau$, 
both 
Gaussian and an exponential decorrelation
model 
(see \S~\ref{sec:statistics_tracer})
can yield values of $D_{\rm corr-fit}$ close to $D_{\rm corr}$. 
However,  it is visually evident that the curve with Gaussian decorrelation provides a better 
description of $A(\tau)$ compared to the exponential decorrelation (see one selected case in 
Figure~\ref{fig:auto-correlation_incomp} of the Appendix~\ref{sec:appendix}).
For the other incompressible 
runs with different values of $M_A$ and resolution, we reach the same conclusion of a 
better description with a Gaussian decorrelation (not shown). This result 
contradicts the assumption of an exponential decorrelation 
(\S~\ref{sec:RD_Theory}). 
Despite this, we find good agreement 
between $D_{\rm corr}$, $D_{\rm corr-fit}$, and $D_{\rm yz} $ for all simulations represented by 
the corresponding symbols on the left panel of Figure~\ref{fig:ma_eta_incomp}.

We notice in the left panel of Figure~\ref{fig:ma_eta_incomp}, however, that the trend of $D_{\perp}$ 
is different for the different resolutions
(compare the solid and dashed lines). The lower resolution simulations seem to be in accordance 
with a power-law dependence $D_{\perp} \propto M_A^{3}$ (excluding the lowest value of $M_A$). 
For the higher resolution simulations, the dependence appears to be closer to the power law 
$D_{\perp} \propto M_A^{2}$. In order to understand what could be affecting the diffusion 
coefficients when we increase the resolution, on the right panel of Figure~\ref{fig:ma_eta_incomp} 
we compare the diffusion coefficient given by $D_{\rm yz}$ (extracted from the spatial displacements 
of the particles) with the average 2D velocity 
(which is composed only by modes with $k_{\parallel} = 0$, 
and takes into account only the solenoidal component (which is the total velocity in the
incompressible case), 
as in the analysis done in~\citealt{santos2021diffusion}). The diffusion provided by these 
2D velocity modes should be similar to their mixing rate, i.e., 
$D_{\rm 2D} \sim \ell_{\rm 2D} \langle v_{\rm 2D, sol} \rangle$, where $\ell_{\rm 2D}$ is the 
dominant scale of the 2D modes (we expect $\ell_{\rm 2D} \lesssim \ell$, the turbulence outer scale). 
Therefore, the normalized curves for $\langle v_{\rm 2D, sol} \rangle$ can give an idea of the 
contribution of $D_{\rm 2D}$ to the total diffusion $D_{\perp}$ (although there is an uncertainty of about 
a factor of unity between $D_{\rm 2D}$ and $\ell \langle v_{\rm 2D, sol} \rangle$).
We can see 
that for the higher resolution simulations (connected by solid lines on the right panel in 
Figure~\ref{fig:ma_eta_incomp}), the contribution of $D_{\rm 2D}$ can be important and 
even dominant for the two smallest values of $M_A$ (observe that the normalized curve of 
$\langle v_{\rm 2D, sol} \rangle$ is above the normalized curve of $D_{\perp}$ for these values 
of $M_A$). We can infer, therefore, that the 2D modes are important for the diffusion for 
$M_A \lesssim 0.2$. 

We 
argue that the transport due to the 2D modes 
is an effect of our numerical setup: a
combination between the limited domain size and the periodic boundary conditions. 
As these modes do not bend the magnetic field lines, there is no suppression in their transport rate, 
as it is expected for the ``true’’ 3D modes. For the lower resolution simulations (dashed lines 
on the right panel of Figure~\ref{fig:auto-correlation_incomp}), we observe a lower level in 
$\langle v_{\rm 2D, sol} \rangle$ compared to the higher resolution simulations. This occurs because the higher numerical 
dissipation in the lower resolution simulations is probably avoiding the growth of the 2D modes 
to 
levels 
that make them important 
for the diffusive transport. In conclusion, in order to observe the dependence of the diffusion coefficient 
on $M_A$ in each resolution, 
only the two highest values of $M_A$ 
($M_A \gtrsim 0.3$) for the simulations with higher resolution, and the three highest values of $M_A$ 
($M_A \gtrsim 0.2$) for the simulations with the lower resolution
should be considered. The fitting of the index $\alpha$ 
in the relation $D_{\rm yz} \propto M_A^{\alpha}$ for these selected simulations in the different 
resolutions is shown in Figure~\ref{fig:ma_eta_ms} (see the gray dotted line for the curve $M_{S} = 0$).

\begin{figure}
\begin{tabular}{c}
        \input{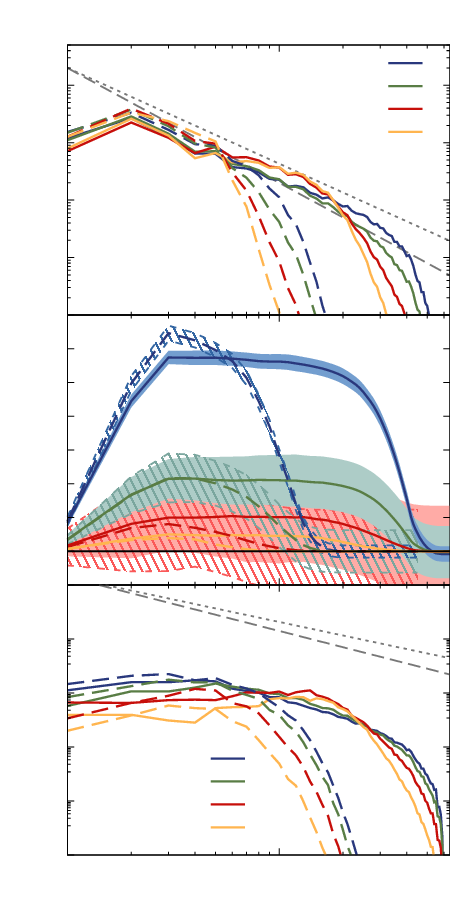} \\
\end{tabular}
\caption{The energy spectrum $E_{\rm 1D} (k_\perp)$ (top panel), the energy transfer spectrum 
$T_{\rm 1D} (k_\perp)$ (middle panel), and the energy spectrum for the 2D velocity modes 
($k_{\perp} = 0$) $E_{\mathbf{u}^{\rm 2D}_{\rm sol}} (k_\perp)$ (bottom panel), for the 
incompressible simulations from Table~\ref{tab:simulations_params}.
Simulations with different resolutions are compared: $2048 \times 128^2$ (continuous lines) 
and $1024 \times 64^2$ (dashed lines).
The shaded intervals shown around the curves of the energy transfer spectrum 
(middle panel) represent the statistical standard deviation of the mean in the temporal 
distribution of the transfer spectra. 
 Hatched (solid) shaded areas 
correspond to simulations with lower (higher) resolution. 
 In order not to impair the readability, these areas are not shown for the simulations with $M_A = 0.12$, as they are too large.
}
\label{fig:ps_incomp}
\end{figure}

Figure~\ref{fig:ps_incomp} shows the 1D energy spectra 
$E_{\rm 1D} (k_{\perp})$ (Eq.~\ref{eq:e1d}) (top panel), the 1D transfer spectrum $T_{\rm 1D} (k_{\perp} )$ (Eq.~\ref{eq:t1d})
(middle panel), and the 1D spectrum of the 2D solenoidal modes (Eq.~\ref{eq:spectrum_vel2d}) (bottom panel) 
for the incompressible simulations. Each curve represents a run 
with a different 
value
of $M_A$. The solid curves correspond to the higher resolution simulations, 
while the dashed curves are for the lower resolution simulations.
The reference curves $\propto k^{-5/3}$ and $\propto k^{-2}$ are shown next to the energy spectra 
for comparison 
as dotted and dashed gray curves, respectively. 
For the transfer spectrum only (middle panel) we present the statistical standard deviation of the mean 
(the mean is taken over snapshots from different times of the simulation, as explained in 
\S~\ref{sec:tracer_methods}) as shaded intervals around the curves. 
For the higher resolution simulations the spectra 
appear
to be flatter than the $\propto k^{-5/3}$ curve, although the inertial range is very 
restricted. The inertial interval can be estimated by the plateaus in the transfer spectrum, 
normalized by $E_{\rm turb} / (\ell / v_{\rm turb})$. We see the suppression of the energy transfer rate 
with decreasing values of $M_A$, as expected by the weak turbulence regime. The energy spectrum 
of the 2D solenoidal velocity modes (lower panel), normalized by the total turbulence energy, shows that the energy 
in these modes is always much 
lower than the energy of the total spectrum (compare the vertical 
scale between the upper and lower panels). It is also notable that the spectra of the 2D modes do 
not peak close to the injection wavenumbers ($k_{\perp, {\rm inj}} L / 2 \pi \approx 2-3$), but at slightly 
higher values (smaller scales). 
Observing the curves of the higher resolution simulations, 
we see that in the two simulations with lower of $M_A$, the peak of this spectrum tends 
to 
shift
to higher $k_{\perp}$ values compared to the simulations with higher values of $M_A$. 
For scales larger than this peak (smaller $k_{\perp}$), the spectrum tends to be flat. 
From comparing the spectra between different resolutions,
it is also clear the increased
importance of the 2D modes for the two high-resolution simulations with the lowest values of 
$M_A$. 
This result is in line
with the analysis of the increased contribution of 2D modes 
to the diffusion coefficient in the higher resolution simulations with the lowest values of $M_A$, 
as compared to the lower resolution simulations.

\subsection{Diffusion coefficient for the compressible simulations}

We also apply the analysis used in the incompressible case to the compressible 
simulations in Table~\ref{tab:simulations_params}.The sonic regimes of the 
compressible simulations are characterized by three values of the sonic Mach number:
$M_S = 0.1$ (subsonic), $M_S = 1$ (transonic), and $M_S = 3$ (supersonic). We aim to 
understand the impact of 
$M_S$ on the properties of the diffusion coefficient 
$D_{\perp}$ of the particles/magnetic field.

\begin{figure*}
\begin{tabular}{c}
        \input{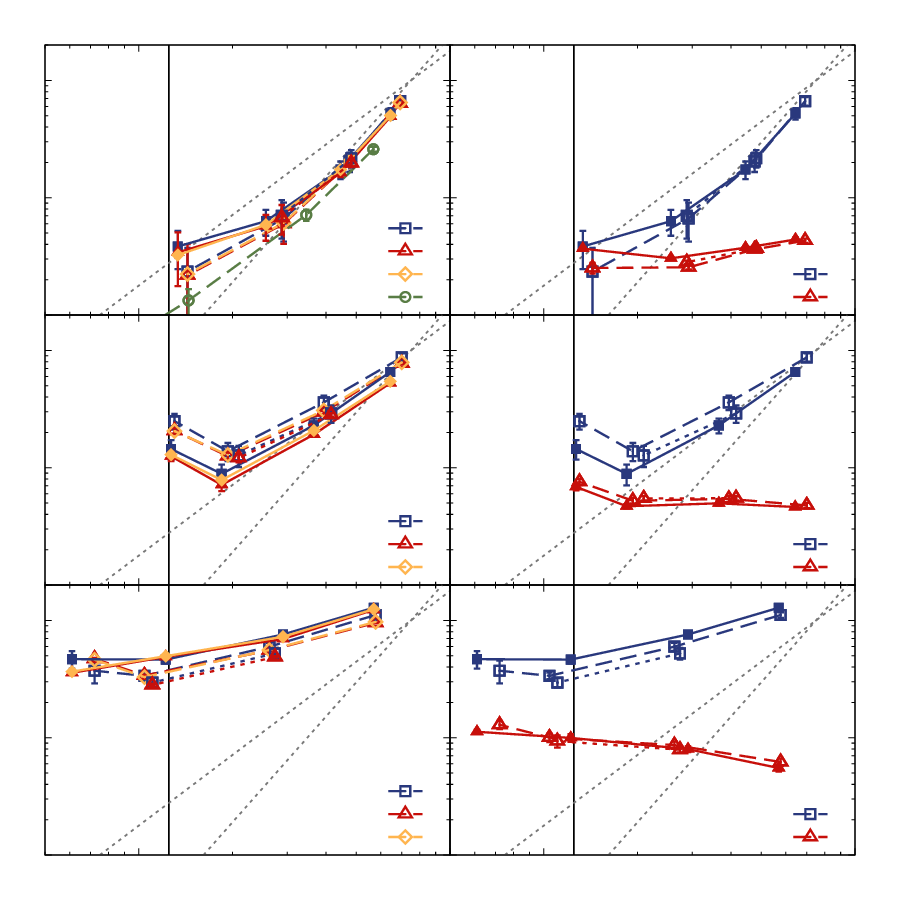}
\end{tabular}
\caption{Normalized diffusion coefficient of the tracer particles $D_{\perp}$ (left and right 
columns), and the rms value of the 2D component of the solenoidal velocity
$\langle v_{2D, {\rm sol}} \rangle$ normalized by the total rms velocity $v_{\rm rms}$ (right only), 
as a function of the Alfvénic Mach number $M_A = v_{\rm rms}/ \langle v_{A} \rangle$. 
Each point on the curves represents a compressible simulation from the run set conducted
using the \textsc{Pluto} code (see Table~\ref{tab:simulations_params}). The only exceptions
are the points labeled $\eta_{\rm tf}$ (Pencil) in the top left panel, which correspond 
to the diffusion coefficient measured with the \textit{Test-Field} method for the run 
set performed with the \textsc{Pencil} code, as shown in Figure~\ref{fig:ma_eta_codes}.
The vertical bars indicate the standard deviation in the temporal distribution 
of the quantities $D_{\rm corr}$, $D_{\rm yz}$, and $\langle v_{2D, {\rm sol}} \rangle$, 
and do not represent statistical errors of the mean values.
Top row: sonic Mach number $M_S=0.1$. Middle row: sonic Mach number $M_S=1$. Bottom row: sonic Mach number $M_S=3$.
The diffusion coefficients are measured using the velocity auto-correlation function 
($D_{\rm corr}$), the evolution of the particles displacements ($D_{\rm yz}$), 
or 
the fitted curve for the velocity auto-correlation function 
($D_{\rm corr-fit}$). 
The vertical line indicates the theoretical limit of $M_A$ below which finite domain size 
effects can affect the turbulence regime (see text for more details). 
Simulations with different resolutions are compared: $2048 \times 128^2$ (continuous lines) 
and $1024 \times 64^2$ 
( long-dashed lines). 
The points connected with 
 short-dashed
lines correspond 
to the runs with names finishing in `CT', having resolution $1024 \times 64^2$ (see text for more details).}
\label{fig:ma_eta_compres}
\end{figure*}

Figure~\ref{fig:ma_eta_compres} shows the diffusion coefficients $D_{\rm corr}$, $D_{\rm corr-fit}$, 
and $D_{\rm yz}$ for the simulations with $M_S = 0.1$, $1$, and $3$ (top, middle, and 
bottom panels, respectively). In the left column, we compare the three diffusion coefficients 
for each simulation, with the solid curves connecting the points representing the 
higher resolution simulations, and the 
 long-dashed 
curves connecting the points for the lower 
resolution simulations. 
The 
 short-dashed 
lines connect the points representing the simulations using the 
constrained transport method for the control of magnetic field divergence (runs with `CT' in 
the final part of the name in Table~\ref{tab:simulations_params}).
We obtain an excellent correspondence between $D_{\rm corr}$ 
and $D_{\rm yz}$ 
in all simulations. Besides, the values 
of the diffusion coefficients for the
runs with constrained transport are nearly indistinguishable from those of the equivalent run using the
divergence cleaning method.
The diffusion coefficients obtained from the fitting of the auto-correlation 
function of the velocity $D_{\rm corr-fit}$ are also in agreement with the other diffusion 
coefficients.
For the subsonic case ($M_S = 0.1$, top left panel), the diffusion coefficients from the 
\textsc{Pencil} runs, measured using the \textit{Test-Field} method (points connected by green line, 
also shown in Figure~\ref{fig:ma_eta_codes}) are shown for comparison. The consistency between 
the results of the different codes reduces the chances that numerical artifacts 
are playing an important role in the diffusivities (at least for the $M_S = 0.1$ case).

By visually inspecting the fitted curves for the autocorrelation $A(\tau)$ and its 
derived cumulant for the compressible models, using the different decorrelation 
models (Gaussian and exponential), it is possible to identify that the best approximation 
for $A(\tau)$ is given by the fit with the Gaussian decorrelation in all subsonic 
($M_S = 0.1$) and most of transonic ($M_S = 1$) simulations 
(a few selected cases are shown in Figure~\ref{fig:auto-correlation_compres} 
of the Appendix~\ref{sec:appendix}).
For the lowest values of $M_A$,
exponential decorrelation gives a better approximation 
(not shown). 
For all supersonic simulations ($M_S = 3$), 
the exponential decorrelation 
is observed to describe $A(\tau)$ better than the Gaussian one.  
In all analysis below, involving the fitted curves of $A(\tau)$ (and their parameters), we use the 
decorrelation model which 
gives the best approximation, as described above.

On the right hand side of Figure~\ref{fig:ma_eta_compres}, the normalized $D_{\rm yz}$ diffusion 
coefficients are compared with the contribution of the 2D modes $D_{\rm 2D}$, estimated 
(within a factor of order unity)
by the normalized 
values of $\langle v_{\rm 2D, sol} \rangle$. The curves connecting the points representing 
the higher resolution simulations are continuous, and those for the lower resolution simulations 
are dashed. 
In the three panels we observe that the tendency  
of $D_{\rm yz}$  to depend on
$M_A$ changes between the two simulations with the lowest values of $M_A$.
These trends coincide with the increasing contribution of $\langle v_{\rm 2D, sol} \rangle$. 
As in the incompressible case, this behavior suggests that the diffusion by the 2D modes becomes 
non-negligible compared to the diffusion provided by the 3D modes. Considering this 2D diffusion 
as an effect of our numerical setup,
we exclude these simulations from 
Figure~\ref{fig:ma_eta_ms}, which shows a fit for the dependence of $D_{\rm yz}$ on $ M_A$ 
in the power law form $D_{\rm yz} \propto M_A^{\alpha}$, for each value of $M_S$. The left (right) panel 
of Figure~\ref{fig:ma_eta_ms} is for the higher (lower) resolution simulations. 
We observe that the value of $\alpha$ decreases with 
the increase of $M_S$, moving away from 
values between $\sim 2$-$3$ when $M_S = 0.1$. 
In \citet{santos2021diffusion} a similar pattern was observed in the simulations with much lower 
values of $M_S$.

\begin{figure*}
\begin{tabular}{c c}
        \input{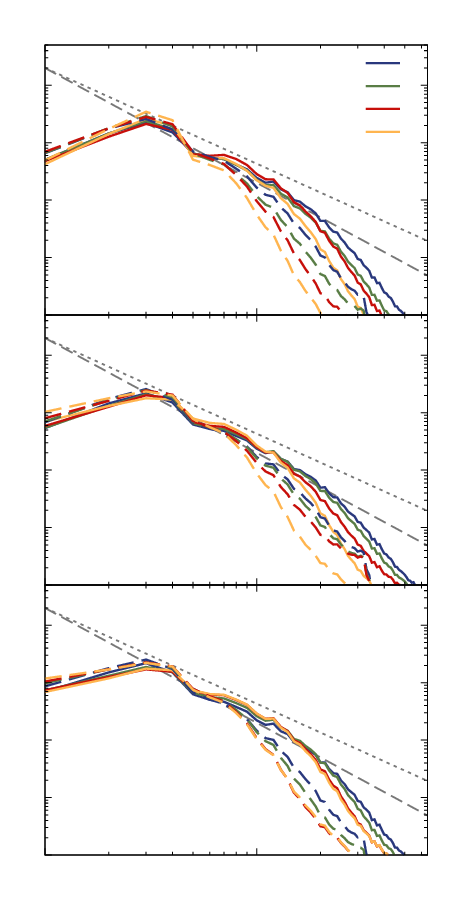} &
        \input{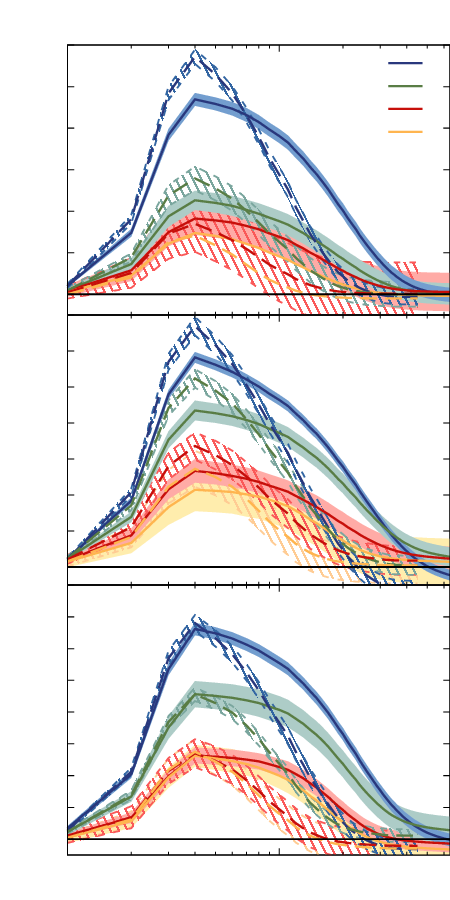}
\end{tabular}
\caption{The energy spectrum $E_{\rm 1D} (k_\perp)$ (left column) and the energy transfer spectrum 
$T_{\rm 1D} (k_\perp)$ (right column), for the compressible simulations from Table~\ref{tab:simulations_params}.
Top row: sonic Mach number $M_S=0.1$. Middle row: sonic Mach number $M_S=1$. Bottom row: sonic Mach number $M_S=3$.
Simulations with different resolutions are compared: $2048 \times 128^2$ (continuous lines) 
and $1024 \times 64^2$ (dashed lines).
The shaded intervals shown around the curves of the energy transfer spectrum 
(left column) represent the statistical standard deviation of the mean in the temporal 
distribution of the transfer spectra. 
Hatched (solid) shaded areas 
correspond to simulations with lower (higher) resolution.
In order not to impair the readability, these areas are not shown for the simulations with $M_A = 0.12$ in the top row ($M_S = 0.1$), as they are too large.
}
\label{fig:ps_compres}
\end{figure*}

Figure~\ref{fig:ps_compres} shows the 1D energy spectra $E_{\rm 1D} (k_{\perp})$ (left column) 
and the transfer spectra $T_{\rm 1D} (k_{\perp})$ (right column) for the compressible simulations.
Each simulation is represented by a curve. The solid curves 
are for higher resolution simulations and the dashed ones are for lower resolution simulations. 
The reference curves $\propto k^{-5/3}$ and $\propto k^{-2}$ are shown alongside the energy 
spectra for comparison.

\begin{figure}
\begin{tabular}{c}
        \input{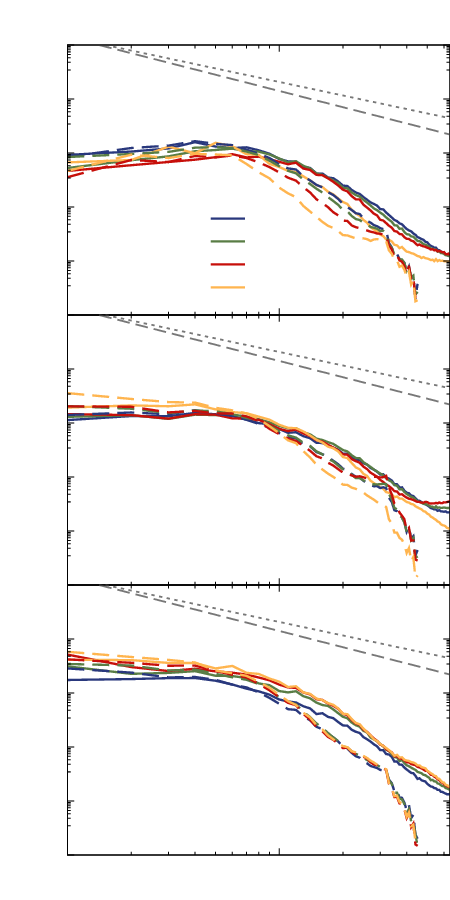}
\end{tabular}
\caption{The energy spectrum for the 2D velocity modes 
($k_{\perp} = 0$) $E_{\mathbf{u}^{\rm 2D}_{\rm sol}} (k_\perp)$, 
for the compressible simulations from Table~\ref{tab:simulations_params}.
Top row: sonic Mach number $M_S=0.1$. Middle row: sonic Mach number $M_S=1$. Bottom row: sonic Mach number $M_S=3$.
Simulations with different resolutions are compared: $2048 \times 128^2$ (continuous lines)
and $1024 \times 64^2$ (dashed lines).}
\label{fig:ps2D_compres}
\end{figure}

We note that some of the transfer spectra from the transonic and supersonic simulations 
($M_S = 1$ and $3$, respectively) do not stabilize at the zero value after the plateau, 
as 
it would be expected. 
This is likely due to
the neglect of the compressible modes in the calculation of 
$T_{\mathbf{k}}$.  
We see the suppression of the energy transfer rate in the perpendicular direction, 
by observing the decreasing of $T_{\rm 1D}$ relative to $E_{\rm turb} / (\ell / v_{\rm rms})$ 
with the reduction of $M_A$. This effect is not as pronounced as in the incompressible case, 
especially for the lowest values of $M_A$. We also observe that the inertial intervals 
(inferred as the region closer to the maximum of the curve $T_{\rm 1D}(k_{\perp})$) are narrower compared 
to those in the incompressible case. This could be expected, as the effective resolution of 
spectral codes tends to be higher than that of second-order finite volume codes.

Figure~\ref{fig:ps2D_compres} exhibits the 1D energy spectra of the 2D solenoidal velocity 
normalized by $E_{\rm turb}$, for the same simulations presented in Figure~\ref{fig:ps_compres}. 
These spectra are always well below the total energy spectra (compare the values in the vertical axes
of both figures). However, the energy of these 2D modes 
increases with resolution, especially at smaller scales (larger $k_{\perp}$) although this effect 
is smaller compared to the incompressible case (probably due to the lower effective resolution 
of the compressible simulations). The shift of the peak in the spectra $E_{\mathbf{u}^{\rm 2D}_{\rm sol}} (k_\perp)$ 
to higher values of $k_{\perp}$ with increasing resolution is most clearly seen for the 
subsonic simulations ($M_S = 0.1$, top panel), following more closely the behavior of the 
incompressible simulations.

\subsection{Dependence of the diffusion coefficient on $M_S$}

\begin{figure*}
\begin{tabular}{c}
        \input{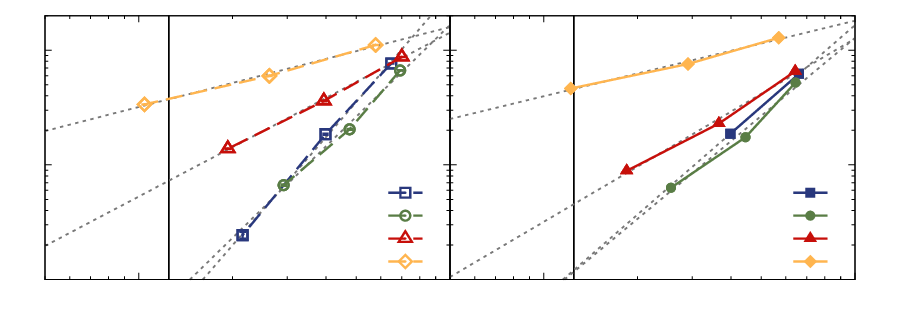}
\end{tabular}
\caption{Normalized diffusion coefficient for the tracer particles $D_{\perp}$ measured by the 
evolution of the particles displacements ($D_{\rm yz}$) 
as a function of the Alfvénic Mach number $M_A = v_{\rm rms}/ \langle v_{A} \rangle$. 
Each curve connects points representing simulations from Table~\ref{tab:simulations_params}
with the same sonic Mach number $M_S$. The simulations are the same presented in 
Figures~\ref{fig:ma_eta_incomp} and~\ref{fig:ma_eta_compres}. Only simulations for which 
the diffusivity by the 2D velocity modes were not interpreted as dominants to the total 
diffusivity are included (see text for more details). 
Left: simulations with resolution $1024 \times 64^2$. Right: simulations with resolution $2048 \times128^2$.
The gray dotted lines are fits to the each curve. 
The vertical bars indicate the statistical error (standard deviation of the mean) 
in the temporal distribution of $D_{\rm yz}$ and are hardly seen due to their amplitude being 
smaller than the sizes of the points. 
The vertical line indicates the theoretical limit of $M_A$ below which finite domain size 
effects can affect the turbulence regime (see text for more details).}
\label{fig:ma_eta_ms}
\end{figure*}

Figure~\ref{fig:ma_eta_ms} shows the dependence of the diffusion coefficients $D_{\rm yz}$ on $M_A$ 
for the incompressible and compressible simulations (each curve connects the points representing 
simulations with the same $M_S$), and a fitting to the parameter $\alpha$ 
at the power-law relation $D_{\rm yz} \propto M_A^{\alpha}$ for each curve. The simulations in the 
left panel have lower resolution, while the simulations in the right panel have higher resolution. 
It is important to highlight that we do not include simulations 
where the contribution 
from the 2D velocity modes to the diffusion was relevant (in general, simulations with 
the lowest values of $M_A$ for each set of simulations).
For each $M_S$, the values of $\alpha$ are 
similar for simulations with different resolutions. This could be expected if the transport is carried out mainly 
by the largest scale motions of the turbulence, close to the injection scale. The increase in 
resolution, however, allows the growth of the 
2D velocity modes in our setup, 
and for some $M_S$ and $M_A$ 
 values,
these modes 
become important for the particle and 
magnetic field transport. 

\begin{figure}
     \input{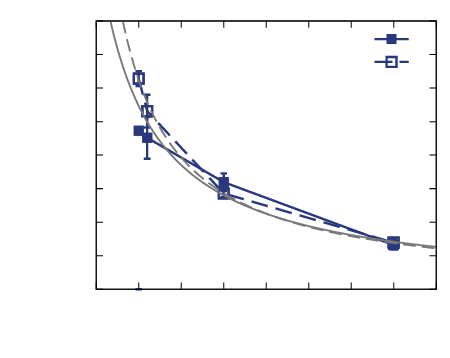} 
\caption{Parameter $\alpha$ as a function of $M_S$, where $\alpha$ is the power law 
in the dependence of the diffusion coefficient on $M_A$ ($D_{\perp} \propto M_A^\alpha$).
Each curve connects the values of $\alpha$ extracted from simulations with the same resolution:
$2048 \times 128^2$ (continuous lines) and $1024 \times 64^2$ (dashed lines).
The vertical bars indicate the error in the fitting values of $\alpha$. 
The gray lines show the fitted curve $\alpha(M_S) = \alpha_0 / (1 + b M_S)$ for each resolution.}
\label{fig:alpha_ms}
\end{figure}

The values of $\alpha$ in the fitted curves with different $M_S$ of Figure~\ref{fig:ma_eta_ms} 
are shown in Figure~\ref{fig:alpha_ms}, 
along with the error in the fit indicated by the vertical bars. 
The solid (dashed) blue line connect the $\alpha$ values for 
the higher (lower) resolution simulations. 
We propose an empirical dependence 
\begin{equation}
\alpha (M_S) = \frac{\alpha_0}{1 + b M_S}, 
\end{equation}
and adjust the values of $\alpha_0$ and $b$ for each resolution. The fitted curves are shown 
in gray color. For the lower resolution, we have 
$\alpha_{0,{\rm lo}}=3.11 \pm 0.04$ and $b_{\rm lo}=1.17 \pm 0.04$,
and for the higher resolution we obtain   
$\alpha_{0,{\rm hi}}=2.73 \pm 0.47$ and $b_{\rm hi}=0.95 \pm 0.29$,
where the fitting procedure accounts for the errors in the $\alpha$ values for each $M_S$.

\subsection{Energy transfer time versus the velocity decorrelation time}

\begin{figure*}
\begin{tabular}{c c}
        \input{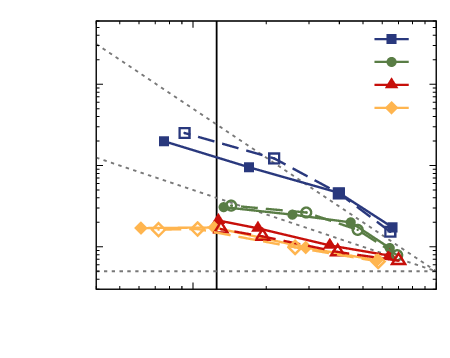} &
        \input{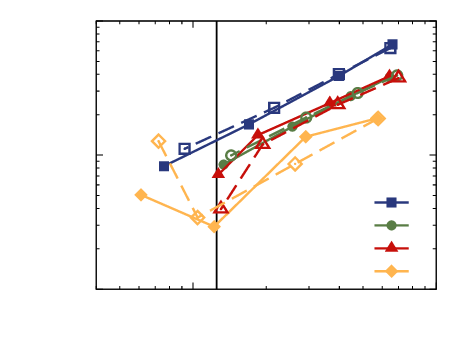} \\
\end{tabular}
\caption{Energy transfer time at the injection scale 
$\tau_{\rm ener} \equiv E_{\rm turb} / T_{\rm turb}$ (left) 
and the velocity decorrelation time $\tau_{\rm dec}$ (right) 
as a function of the Alfvénic Mach number $M_A = v_{\rm rms}/ \langle v_{A} \rangle$. 
Each curve connects points representing simulations from Table~\ref{tab:simulations_params}
with the same sonic Mach number $M_S$. The simulations are the same presented in 
Figures~\ref{fig:ma_eta_incomp} and~\ref{fig:ma_eta_compres}. 
Simulations with different resolutions are compared: $2048 \times 128^2$ (continuous lines) 
and $1024 \times 64^2$ (dashed lines).}
\label{fig:ma_gamma_tener}
\end{figure*}

The left panel of Figure~\ref{fig:ma_gamma_tener} shows, for compressible and incompressible 
simulations, the dependence of the energy transfer time at the injection scale 
$\tau_{\rm ener}$ on $M_A$. The energy transfer time is normalized by the non-linear time 
$\ell / v_{\rm rms}$. Each curve connects the points representing simulations with the same 
value of $M_S$. The solid curves are for the higher resolution simulations and the dashed 
curves are for low-resolution ones. A few reference curves are plotted for comparison: 
$\propto M_A^{0}$, $\propto M_A^{-1}$, and $\propto M_A^{-2}$. The theory of weak Alfvénic 
turbulence predicts a dependence $\tau_{\rm ener} \propto M_A^{-1}$, while in the strong 
turbulence regime we expect no dependence on $M_A$. We note that for the incompressible 
simulations the energy transfer time seems to be close to $\tau_{\rm ener} \propto M_A^{-1}$. 
In the simulations with $M_S = 0.1$, $1$, and $3$, however, the curves are between the dependences 
$M_A^{0}$ and $M_A^{-1}$, apparently closer to $M_A^ {0}$. 

One of the implicit assumptions in recovering the results of the RD theory is the 
relation between the velocity decorrelation time and the energy transfer time 
(see \S~\ref{sec:RD_Theory}).
In the right panel of Figure~\ref{fig:ma_gamma_tener} we show the
characteristic decorrelation time (either exponential or Gaussian, which gave the best fitting to
the velocity auto-correlation function). We observe 
that the decorrelation time 
increases linearly with $M_A$, that is, $\tau_{\rm dec} \sim \ell / V_A$. Therefore, at least for our simulations,
the velocity decorrelation time does not coincide with the energy transfer time at the injection scale.

\section{Discussion}

\subsection{Comparison with previous studies}

In \citet{santos2021diffusion}, simulations of forced MHD turbulence in the sub-Alfvénic 
regime with low compressibility were conducted in periodic domains. The aim was to use the 
Test-Field method to extract the dependence of the magnetic field diffusion coefficients 
on 
$M_A$, and to compare these findings with the predictions of 
reconnection diffusion theory. This previous study analyzed the influence of 2D modes in the forcing
and the size of the computational domain. To promote the development of weak (Alfvénic) 
turbulence and avoid the dominance of 2D modes in the turbulence cascade, a forcing with 
modulated amplitude was employed. This approach 
excluded the 2D modes in the forcing 
and used a domain size in the direction of the uniform magnetic field that, according to 
theoretical predictions based on reduced MHD (see \citealt{nazarenko20072d}), was most suitable.
The dependence 
of the diffusion coefficient $D_{\perp}$ 
on
$M_A$ was shown to be in close agreement with the 
prediction of the RD theory ($D_{\perp} \sim \ell_{\rm turb} v_{\rm rms} M_A^{3}$) for the 
simulations with the lowest values of the sonic Mach number, $M_S = 0.02$. For the simulations 
with higher values of $M_S$ ($=0.08$ and $0.32$), a weaker suppression  of the 
diffusion was observed, close to $D_{\perp} \propto M_A^{2}$. The dependence of the energy cascading time (from the injection 
scale) 
on 
$M_A$ did not coincide with the predictions of the weak regime 
($\tau_{\rm ener} \sim M_A^{-1}$), showing a stronger dependence ($\tau_{\rm ener} \sim M_A^{-2}$) 
for the simulations with the highest values of $M_A$.
For the smallest values of $M_A$ 
this 
dependence weakened to nearly
$\sim M_A^{0}$, 
and the transition limit of $M_A$ for this behavior was shown to 
increase with $M_S$.

In the present study, we investigated in detail the behavior of the diffusion coefficient 
in sub-Alfvénic MHD turbulence, 
for a wide range of sonic 
regimes ($M_S$ = 0, 0.1 and 3). We employed different numerical techniques 
suitable for each of these regimes. We also investigated several of the assumptions 
implicit in the quantitative predictions of the RD theory. To address
the first of these assumptions (see \S2),
we used a set of Lagrangian tracer particles 
that are confined to move only in the direction 
perpendicular to $\mathbf{B}_0$ in the MHD simulations. 
Then, we extracted their perpendicular 
diffusion rates through two different methods: the analysis of the auto-correlation function 
of the velocity in time (which requires a high-frequency time series in the output data, 
in order to capture the highest frequency modes in the simulations), and a direct analysis 
of the spatial displacement of particles (which is computationally less expensive in terms of memory, 
as high-frequency time series are not required). For all our simulations, we obtained 
a good agreement between these measurements. We used the \textsc{Pencil} code to produce a set of 
low compressibility simulations, where in addition to the particles tracer 
we also used 
the Test-Field method, and for this set of simulations we demonstrated the agreement between 
the magnetic field 
and the particles diffusion.
Therefore, we have demonstrated 
the first assumption of the RD theory, and introduced two 
new practical methods that we used in the incompressible and compressible simulations 
to quantify the magnetic flux diffusion coefficient. The validity of this assumption also 
shows that the perpendicular transport rates obtained in this work apply to the transport 
of scalar fields (such as temperature and chemical composition).

The quantification of the transport coefficient in the incompressible case is in close agreement 
with the prediction of RD theory (which 
was derived exactly in the incompressible limit) for the 
suppression of transport with the magnetic field, $\eta_{\rm rd} \propto M_A^{3}$.
However our  results for  the compressible simulations show an 
that the suppression ins mitigated with the increase of 
$M_S$, 
and it seems to point to no suppression in the asymptotic case where $M_S \gg 1$. 
We propose an empirical description for the dependence of $\eta_{\rm rd}$ on $M_A$ and $M_S$, 
in the form $\eta_{\rm rd} \sim \ell v_{\rm rms} M_A^{\alpha_0 / (1 + b M_S)}$. 
A fit for this 
dependence from the analysis of our lower resolution simulations, where the contribution to the diffusion 
from the 2D velocity modes is probably less important, gives $\alpha_0 \approx 3.1$ and $b \approx 1.2$. 
This empirical relation can be applied to the evaluation of the magnetic field diffusion 
rate by the RD in molecular clouds and the diffuse ISM. Around clouds of the ISM, on scales 
$\sim 0.1$ pc, recent studies based on starlight polarization have shown turbulence to be sub-Alfvénic 
(\citealt{versteeg_etal_2024, angarita_etal_2024, yasuo_etal_2024}).
The role of the RD in star 
proto-planetary disk formation has been explored in a few studies in the last decade, 
and these studies show that it can be very effective in comparison to other possible mechanisms 
of magnetic flux transport, such as ohmic resistivity and ambipolar diffusion 
(see for example \citealt{santos2010diffusion, krasnopolsky_etal_2011, 
lazarian2012magnetization, 
santos2012role, santos2013disc, leao2013collapse, gonzalez2016magnetic, lam_etal_2019}). 
The transport 
of magnetic fields through the gas is an essential ingredient in understanding different 
phases of star formation and the formation of proto-planetary disks.

\subsection{Possible limitations of our results: the role of the transport due to the 2D velocity 
modes, forcing effects, and turbulence regime}

It is important to highlight that in this study we considered that contribution to the 
magnetic field/particles diffusion due to the 2D velocity modes is an effect of our 
periodic boundary conditions and limited size numerical setup.
Moreover, we constrained the
simulations in order to exclude the ones for which contribution from the 2D modes 
seemed to be important or dominant. As for these plasma displacements, there is no restoring 
force from the magnetic field, and the transport they provide is not suppressed, that is, 
there is no dependence on the $M_A$, resulting in the diffusion rate given by the hydrodynamic 
mixing rate $D_{ \rm 2D} \sim \ell v_{\rm 2D}$. Our elongated computational domains 
and zero injection of 2D modes do not prevent their natural development from the nonlinear 
interactions in the turbulence. 
A relevant question is whether their transport would also be present in astrophysical systems,
such as molecular clouds. Although these systems are finite in size, their boundary conditions
are not periodic, as the field lines connect with those of the external plasma environment.
In this study we make a strong assumption that the
transport by these modes would not be effective in real systems, capable of transporting 
field lines without reaction from the magnetic tension forces. 
Recently, \citet{lazarian_etal_2025} investigated the role of these 2D 
modes in systems such as molecular clouds, and how they impact the estimates of magnetic 
fields using methods based on the hypothesis of equipartition 
between kinetic and magnetic energy during the analysis of polarization angles dispersion. 

In our higher resolution simulations, transport through 2D modes gained importance compared 
to the lower resolution simulations. We observed this effect to be 
prominent for simulations 
with the lowest values of $M_A$, due to the strong suppression of transport by purely 3D modes 
-- at least in regimes of none or little compressibility. At least for the highest values 
of $M_A$ in our simulations, for all the values of $M_S$ considered,
the total diffusion coefficient 
did not increase with the
resolution, which may indicate a saturation of the contribution 
from the 2D transport.
However, since higher resolution revealed an increased contribution of 2D diffusion for the smallest
values of $M_A$, it is possible that the effect of $D_{\rm 2D}$ has not yet converged.
For the other simulations, 
a marginal effect from $D_{\rm 2D}$ may still be present, to a greater extent in the higher 
resolution simulations. From our hypothesis that this effect is not important in 
real systems, we suggest that diffusion coefficients obtained from the lower resolution 
simulations may be closer to the real physical values. Following this reasoning, the values 
obtained for the magnetic field diffusion represent upper limits, as there may still be a 
contribution from the 2D modes.

The weak dependence on
$M_A$ for the diffusion in our supersonic simulations 
suggests a
possible asymptotic independence for $M_S \gg 1$, similar to the transport rate observed in
hydrodynamic turbulence. Naively, an asymptotic transport rate corresponding to that 
predicted by strong turbulence could be expected, with 
$\tau_{\rm dec} \sim \tau_{\rm ener} \sim \ell / v_{\rm rms}$, which would result in 
$\eta \sim \ell v_{\rm rms} M_A^{2}$ 
(see \S~\ref{sec:RD_Theory}).
In our transonic and supersonic simulations ($M_S = 1$ and $3$), 
the dependence of $\tau_{\rm ener}$ at the injection scale on $M_A$ is quite weak, 
although not completely independent (between $M_A^{0}$ and $M_A^{-1}$). The physical reason 
for this weak dependence must still be investigated. 
Another point of caution is the validity of the correspondence between particle perpendicular 
diffusion and magnetic field diffusion in the supersonic regime. We demonstrated this
correspondence using weakly compressible simulations, where the Test-Field analysis in the
\textsc{Pencil Code} are straightforward. Although there is no obvious reason why this correspondence 
would not hold for highly compressible simulations, it is essential to verify this in future studies.

Some implicit assumptions in the RD theory did not prove to be valid in the results of our simulations. 
First, 
 regarding the second assumption (see \S2),
we observed that a faster, Gaussian 
time decorrelation provides a better description of the auto-correlation function of the particles 
velocities, even in the incompressible case. The only exceptions were the supersonic 
($M_S = 3$) and in the transonic ($M_S = 1$) simulations with the lowest value of $M_A$ 
($M_A = 0.12$). Second, regarding the third assumption, the decorrelation time scale does not exhibit 
the same behavior as the energy transfer time at the injection scale (compare the left and right 
panels of Figure~\ref{fig:ma_gamma_tener}).
The time scales $\tau_{\rm dec}$ in our simulations follow the linear wave crossing time.
It is possible that this is an effect of our forcing scheme, 
approximately delta correlated in time. 

Furthermore, as 
the diffusion depends on the properties of plasma 
motions
at the injection scale, it is possible 
that our results (including the prediction of the RD for the incompressible limit) are not 
``universal''. 
Only a study dedicated to the influence of the forcing properties on $\eta_{\rm rd}$ 
can elucidate this issue.

Finally, our incompressible simulations show a dependence on
the energy cascading time with $M_A$ 
very close to that predicted by the weak Alfvénic turbulence theory, $\tau_{\rm ener} \propto M_A^{-1}$ 
(Galtier et al. 2000). Although with a very limited inertial range, 
and a spectrum flatter than that predicted by theory ($\propto k_{\perp}^{-2}$), it is possible 
that our incompressible simulations are close to the theoretical regime of weak turbulence. 
The identification of this regime was reported before in simulations of decaying turbulence 
(\citealt{meyrand_etal_2015}) 
and less clearly in forced turbulence in reduced MHD (\citealt{perez_boldyrev_2008}).

\section{Summary}

Below we summarize the main findings of this work:

\begin{enumerate}
\item We present two new techniques to measure the transverse diffusion of magnetic fields, 
using the velocity statistics of several thousands of tracer particles confined to move in the 
plane perpendicular to the uniform magnetic field. The first method is based on the 
calculation of the time correlation of the particles velocities. The second method uses the 
statistical analysis of the displacement of the particles. We show that the diffusion coefficient 
extracted using both methods coincide with the one accessed with the theoretically more 
sophisticated and already well studied \textit{Test-Field} method, based on the analysis 
of the transport of passive vector fields. This correspondence confirms the theoretical 
assumption that the magnetic field diffusion happens at the same rate as the diffusion of 
fluid Lagrangian particles, and at the same time justifies the application of the RD theory 
to the turbulent transport of scalar fields, such as temperature, chemical composition,
and advection of cosmic rays.

\item The analysis of the tracer particles velocity statistics shows that the temporal 
correlation of the turbulence velocity fluctuations decays with a Gaussian dependence 
in time in the subsonic and transonic simulations. For supersonic simulations the temporal 
decorrelation is exponential. 
Further investigations of the effects of the injection turbulence properties on
the temporal correlation should be carried out for advancing the development of the RD theory.

\item The 2D velocity modes ($k_{\parallel} = 0$) develop in our periodic simulations 
and become important for the particles and magnetic field diffusion for the simulations 
with higher resolution and smaller values of $M_A$. We consider this an effect of our 
setup and do not include the effective diffusion resulting from these simulations in the 
assessment of the dependence of the diffusion coefficient on the turbulence parameters.
      
\item Simulations in the incompressible limit are consistent with the predictions of the 
RD theory for the dependence of the diffusion coefficient $D \propto M_A^3$. This is in agreement 
with evidences pointed out in the turbulence regime characterized by low sonic Mach numbers 
(\citealt{santos2021diffusion}).
      
\item Our transonic and supersonic MHD simulations show that the predicted suppression 
of the diffusion coefficient in the presence of strong magnetic fields is reduced with 
the increase of $M_S$. We extract a dependence which can be described by 
$\eta \sim \ell v_{\rm rms} M_A^{\alpha_0 / (1 + b M_S)}$, with $\alpha_0 \approx 3.1$ 
and $b \approx 1.2$. This implies that in star-forming regions (molecular clouds)
the efficiency of turbulent transport of magnetic fields is substantially higher than initially predicted by the RD theory
(based on incompressible Alfvénic turbulence).
\end{enumerate}

Our present results provide strong evidence for the validity of the RD predictions when 
turbulence is close to the incompressible regime, and therefore in the same scenario invoked 
by the theory, and shows how the diffusion coefficient behaves under the more general 
compressible MHD turbulence regime.

The quantitative characterization of the effective diffusion coefficient of the magnetic 
field is critical for the modelling of the star formation process in turbulent molecular 
clouds, and for the evaluation of this transport process efficiency when compared to other 
mechanisms, for instance the ambipolar diffusion and the ohmic diffusion.

\section*{Acknowledgements}

CNK acknowledges support from a PROEX grant of the Brazilian Agency CAPES (88887.684501/2022-00) 
and a grant of the Brazilian Agency FAPESP (2022/10102-4). MVdV is supported by the Grants 2019/05757-9 and 2020/08729-3, Fundação de Amparo à Pesquisa do Estado de São Paulo (FAPESP). 
EMdGDP also acknowledges support from the FAPESP grants (2013/10559-5 and 2021-02120-0) and the National Reseach Council CNPq (grant 308643/2017-8).
Part of the numerical simulations presented here were performed in the \textit{syrtari} server, acquired by the FAPESP grant 2019/05757-9. 
This work also made use of the computing facilities of the Laboratory of Astroinformatics 
(IAG/USP, NAT/Unicsul), whose purchase was made possible also by FAPESP (grant 2009/54006-4).
The authors are indebted to the anonymous reviewer, who helped improve the 
quality and clarity of this work with their numerous suggestions, insightful comments, 
corrections, and for pointing out the caution needed when using the divergence cleaning method when 
flows are highly nonlinear.

\section*{Data Availability}


The data underlying this article will be shared on reasonable request to the corresponding author.



\bibliographystyle{mnras}
\bibliography{camila-diffusion} 




\appendix



\section{Velocity autocorrelation}\label{sec:appendix}

\begin{figure*}
    \begin{tabular} {c c}
         \input{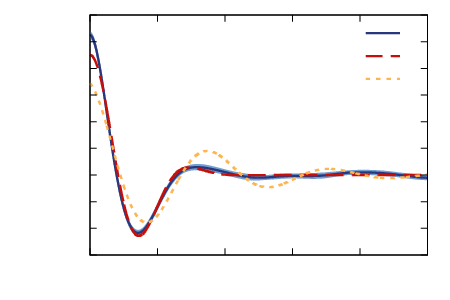} &
         \input{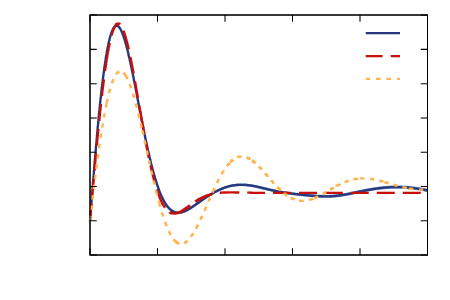}
    \end{tabular}
\caption{Auto-correlation function of the particles velocity (left panel) and its cumulant (right panel) 
as a function of the time interval $\tau$ for the incompressible simulation 
incomp\_ma0.5\_hi\_s 
(see Table~\ref{tab:simulations_params}). The fitted curves for the auto-correlation and its derived 
cumulant are shown, for different decorrelation models: Gaussian (label `fit gauss') 
and exponential (label `fit exp').
The shaded intervals shown around the autocorrelation and its cumulant (blue lines) 
represent the statistical standard deviation in the temporal distribution used for the averaging 
of these quantities at each $\tau$. 
}
\label{fig:auto-correlation_incomp}
\end{figure*}

\begin{figure*}
\begin{tabular} {c c}
         \input{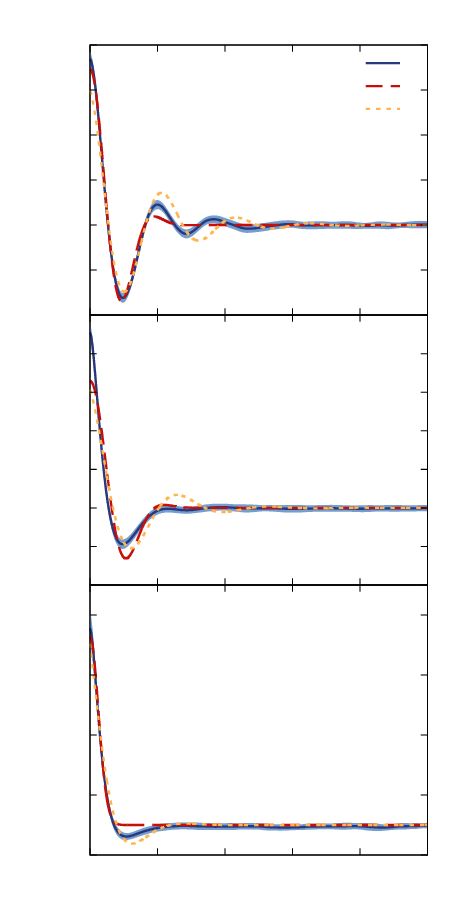} &
         \input{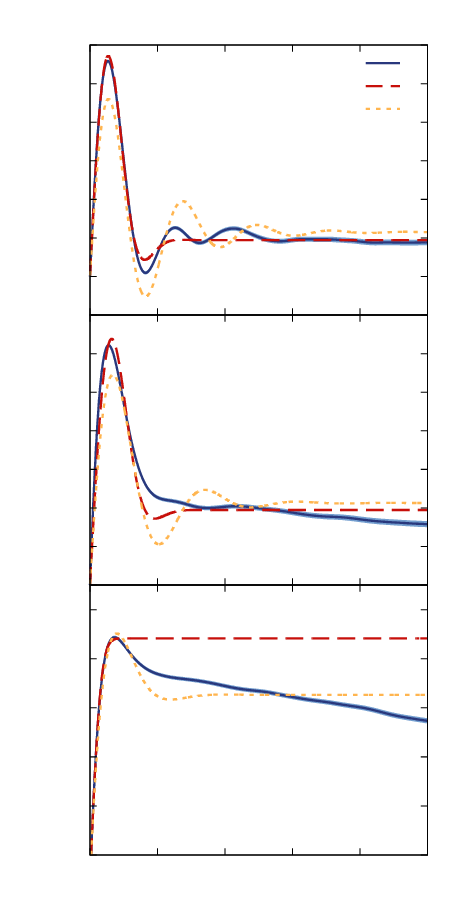} 
\end{tabular}
\caption{Same as Figure~\ref{fig:auto-correlation_incomp} for the compressible simulations 
ms0.1\_ma0.5\_hi\_pl (top row), 
ms1\_ma0.5\_hi\_pl (middle row), and 
ms3\_ma0.5\_hi\_pl (bottom row) 
(see Table~\ref{tab:simulations_params}).}
\label{fig:auto-correlation_compres}
\end{figure*}

The velocity auto-correlation function and its cumulant for one incompressible run 
from Table~\ref{tab:simulations_params}
(incomp\_ma0.5\_hi\_s) are shown on the left and right panels of 
Figure~\ref{fig:auto-correlation_incomp}, respectively. 
The same quantities for a few compressible runs from Table~\ref{tab:simulations_params}
are shown in Figure~\ref{fig:auto-correlation_compres}: 
ms0.1\_ma0.5\_hi\_pl (top panels), ms1\_ma0.5\_hi\_pl (middle panels), and ms3\_ma0.5\_hi\_pl
(bottom panels). 
The fitted curves following either a Gaussian or 
an exponential decorrelation are also shown in Figures~\ref{fig:auto-correlation_incomp} 
and~\ref{fig:auto-correlation_compres}
(see \S~\ref{sec:statistics_tracer}). 

\bsp	
\label{lastpage}
\end{document}